\documentclass[fleqn,10pt]{wlscirep}
\usepackage[utf8]{inputenc}
\usepackage[T1]{fontenc}
\usepackage{graphicx}
\usepackage{amsmath}
\usepackage{braket}
\usepackage{braket}
\usepackage{cite}
\usepackage[utf8]{inputenc}
\usepackage{amssymb}
\usepackage{amsmath}
\usepackage{hyperref}
\usepackage{mathdots}
\usepackage{placeins}

\usepackage{bm}
\usepackage{subfigure}
\usepackage{pgfplots}
\pgfplotsset{compat=1.18}
\usepgfplotslibrary{groupplots}
\usepackage{multirow}
\usepackage{color}
\usepackage{url}
\usepackage{comment}
\usepackage[normalem]{ulem} %to strike the words
\DeclareMathOperator*{\argmin}{argmin}

\usepackage{cases}
\usepackage{cleveref}
\usepackage[acronym]{glossaries}
\glsdisablehyper
\usepackage{layouts}
\usepackage{lineno}
\usepackage{todonotes}

\newacronym{ae}{AE}{annealing engine}
\newacronym{da}{DA}{digital annealing}
\newacronym{fem}{FEM}{finite element method}
\newacronym{qa}{QA}{quantum annealing}
\newacronym{qubo}{QUBO}{quadratic unconstrained optimization}

\newcommand{\lp}{\left(}
\newcommand{\rp}{\right)}

\newcommand{\domain}{\Omega}
\newcommand{\domainBoundaryDirichlet}{\partial\domain_D}
\newcommand{\domainBoundaryNeumann}{\partial\domain_N}
\newcommand{\normal}{\bm{n}}
\newcommand{\fluid}{f}
\newcommand{\domainFluid}{\domain_{\fluid}}
\newcommand{\solid}{s}
\newcommand{\domainSolid}{\domain_{\solid}}
\newcommand{\interface}{\partial\Omega}
\newcommand{\coord}{\bm{x}}
\newcommand{\charFunc}[1][]{\chi_{#1}}

\newcommand{\resistanceCoeff}{\alpha}
\newcommand{\resistanceCoeffSolid}{\bar{\resistanceCoeff}}
\newcommand{\levelSet}{\phi}

\newcommand{\velocity}{\bm{u}}
\newcommand{\pressure}{p}
\newcommand{\strainRateTensor}{\varepsilon}
\newcommand{\viscosity}{\mu}
\newcommand{\objective}{J}
\newcommand{\objectiveRegularized}{\bar{\objective}}
\newcommand{\regularization}{R}
\newcommand{\regularizationCoeff}{\tau}

\newcommand{\volumeFluid}{V_{\fluid}}
\newcommand{\volumeLimit}{V_{\text{max}}}
\newcommand{\elemIndex}{k}
\newcommand{\levelSetElem}{\levelSet^{\elemIndex}}
\newcommand{\levelSetElemNew}{\hat{\levelSet}^{\elemIndex}}
\newcommand{\charFuncElem}[1][]{\chi^{\elemIndex}_{#1}}
\newcommand{\charFuncElemNew}[1][]{\hat{\chi}^{\elemIndex}_{#1}}
\newcommand{\objectiveQUBO}{H}
\newcommand{\objectiveDissipation}{H_{\text{dis}}}
\newcommand{\paramDissipation}{\lambda_{\text{dis}}}
\newcommand{\objectiveRegularization}{H_{\text{reg}}}
\newcommand{\paramRegularization}{\lambda_{\text{reg}}}
\newcommand{\objectiveVolume}{H_{\text{vol}}}
\newcommand{\objectiveQUBOCondensed}{\hat{H}}
\newcommand{\paramVolume}{\lambda_{\text{vol}}}
\newcommand{\objectiveCharacteristic}{H_{\text{char}}}
\newcommand{\paramCharacteristic}{\lambda_{\text{char}}}
\newcommand{\timeOut}{T_{\text{out}}}
\newcommand{\nopt}{n}
\title{An Ising Machine Formulation for Design Updates in Topology Optimization of Flow Channels}

\author[1]{Yudai Suzuki}
\author[2]{Shiori Aoki}
\author[3,*]{Fabian Key}
\author[4]{Katsuhiro Endo}
\author[5,6]{Yoshiki Matsuda}
\author[7,1,6,8,9]{Shu Tanaka}
\author[10]{Marek Behr}
\author[11,1,9]{Mayu Muramatsu}
\affil[1]{Quantum Computing Center, Keio University, Yokohama, Kanagawa, Japan}
\affil[2]{Graduate School of Science and Technology, Keio University, Yokohama, Kanagawa, Japan}
\affil[3]{Institute of Lightweight Design and Structural Biomechanics, TU Wien, Austria}
\affil[4]{Research Center for Computational Design of Advanced Functional Materials, National Institute of Advanced Industrial
Science and Technology (AIST), Tsukuba, Ibaraki, Japan}
\affil[5]{Fixstars Corporation, Tokyo, Japan}
\affil[6]{Green Computing System Research Organization, Waseda University, Tokyo, Japan}
\affil[7]{Department of Applied Physics and Physico-Informatics, Keio University, Yokohama, Kanagawa, Japan}
\affil[8]{Human Biology-Microbiome-Quantum Research Center (WPI-Bio2Q), Keio University, Tokyo, Japan}
\affil[9]{Keio University Sustainable Quantum Artificial Intelligence Center (KSQAIC), Keio University, Tokyo, Japan}
\affil[10]{Chair for Computational Analysis of Technical Systems, RWTH Aachen University, Germany}
\affil[11]{Department of Mechanical Engineering, Keio University, Yokohama, Kanagawa, Japan}

\affil[*]{email: fabian.key@tuwien.ac.at}

\begin{abstract}
Topology optimization is an essential tool in computational engineering, for example, to improve the design and efficiency of flow channels. At the same time, Ising machines, including digital or quantum annealers, have been used as efficient solvers for combinatorial optimization problems. Beyond combinatorial optimization, recent works have demonstrated applicability to other engineering tasks by tailoring corresponding problem formulations. In this study, we present a novel Ising machine formulation for computing design updates during topology optimization with the goal of minimizing dissipation energy in flow channels. We explore the potential of this approach to improve the efficiency and performance of the optimization process. To this end, we conduct experiments to study the impact of various factors within the novel formulation. Additionally, we compare it to a classical method using the number of optimization steps and the final values of the objective function as indicators of the time intensity of the optimization and the performance of the resulting designs, respectively. Our findings show that the proposed update strategy can accelerate the topology optimization process while producing comparable designs. However, it tends to be less exploratory, which may lead to lower performance of the designs. These results highlight the potential of incorporating Ising formulations for optimization tasks but also show their limitations when used to compute design updates in an iterative optimization process. In conclusion, this work provides an efficient alternative for design updates in topology optimization and enhances the understanding of integrating Ising machine formulations in engineering optimization.

\end{abstract}
\begin{document}

\flushbottom
\maketitle
\thispagestyle{empty}

% TEXT WIDTH IN PT
% \printinunitsof{pt}\prntlen{\textwidth}
%%%%%%%%%%%%%%%%%%
%% INTRODUCTION %%
%%%%%%%%%%%%%%%%%%
\section*{Introduction}
\label{sec:introduction}
%% BACKGROUND AND STATE OF THE ART
Ongoing rapid advances in the implementation of \textit{Ising machines}~\cite{Mohseni2022}, which are specialized computing hardware designed to solve combinatorial optimization problems, make it worthwhile to study their application to engineering optimization problems, promising superior efficiency for complex problem instances.
A prominent example of an Ising machine is \textit{\gls{qa}}, which effectively utilizes quantum-mechanical effects such as quantum fluctuations for solving optimization tasks~\cite{Apolloni1989,Finnila1994, Kadowaki1998}.
Since \textit{D-Wave Systems} made \gls{qa} devices available for commercial use, numerous efforts have been made to develop this technology from both software and hardware perspectives~\cite{Tanaka2017,Albash2018,Hauke2020,Yarkoni2022}.
Also, the development of classical and quantum-inspired implementations is progressing quickly~\cite{Inagaki2016,Goto2019,Matsubara2020,Goto2021,Tatsumura2021}.
With these technologies, today's focus is on whether approaches using Ising machines will prove advantageous over conventional methods in engineering optimization.
\par
A key requirement for the development of new optimization approaches using Ising machines is to express the given optimization problem or subproblems in the form of an Ising model. This model involves spins that can point either up or down, which can be described using binary-valued \textit{spin variables} $s_i\in\{-1,1\}$, where $s_i=+1$ for ``up'' and $s_i=-1$ for ``down''. Each configuration of the spin variables is related to the energy state of the model by the so-called \textit{Ising Hamiltonian}, a function of the spin variables $s_i$.
Finding the minimum of this function, i.e., the lowest energy state of the Ising model, is then the task of the Ising machine, which provides the solution to the optimization problem. 
Note that the Ising model can be transformed into a \textit{\gls{qubo} problem} by the change of variables $s_i = 1 -2q_i$, with \textit{binary variables} $q_i\in\{0,1\}$.
Consequently, both problems can be regarded as equivalent, and any problem given in \gls{qubo} form can also be handled by Ising machines.
Nevertheless, binary variables $q_i$ that take values of 0 or 1 are often more intuitive and directly compatible with many real-world design optimization problems that involve decision making (e.g., material/no material, include/exclude).
So, the remaining task will be the construction of a corresponding \gls{qubo} formulation for the original problem of interest.
\bigskip\par
The main area in which Ising machines are expected to have an advantage over conventional computers is combinatorial and discrete optimization problems. From this perspective, many studies have demonstrated the usefulness of Ising machines for relevant combinatorial optimization tasks such as traffic flow optimization~\cite{Neukart2017,Ohzeki2019,Stollenwerk2019, Kanai2024}, scheduling~\cite{Rieffel2015} and portfolio optimization~\cite{Rosenberg2016,Venturelli2019}.
\par
In addition, recent works have proposed applications of Ising machines to engineering tasks, using the tailored \gls{qubo} formulations for problems that were not originally combinatorial or discrete optimization problems.
For instance, \gls{qa} can be used to solve problems arising from the \gls{fem} by deriving a \gls{qubo} problem to minimize the residual in the resulting linear system of equations~\cite{Vreumingen2019,Raisuddin2022}.
Additional works exist, for example, for the simulations of diblock polymers using phase-field models~\cite{Endo2022} or fluid dynamics simulations~\cite{Ray2022}.
\par
Further publications presented the application of Ising machines in the field of design optimization, including topology optimization.
For example, a quantum-classical hybrid methodology was developed for solving continuum topology optimization problems formulated as mixed-integer nonlinear programs~\cite{Ye2023}. This approach uses a splitting method where two types of subproblems are solved, one on a classical computer and the other by \gls{qa}.
In another work, a novel method for topology optimization of truss structures using \gls{qa} in an iterative scheme has been developed, where both state and design variables are determined by solving a \gls{qubo} problem~\cite{Honda2024}. The optimization of truss structures has also been considered using a nested algorithm in which the truss system is analyzed using the \gls{fem} and design updates for the cross sections are determined by \gls{qa}~\cite{Wang2024}.
In addition, the application of topology optimization via \gls{da} to the design of electromagnetic devices was studied as well~\cite{Maruo2022}.
In distinction to the previous works, a formulation for structural design optimization, which is specifically designed for Ising machines and does not rely on iterations or classical methods such as the \gls{fem}, was applied to a test case involving a rod under self-weight loading~\cite{Key2024}.
Additional results were presented for applications such as the design of electric circuits~\cite{Okada2023}, photonic-crystal surface-emitting lasers~\cite{Inoue2022}, printed circuit boards~\cite{Matsumori2022}, wavelength selective radiators~\cite{Kitai2020}, barrier materials~\cite{Nawa2023}, or multimolecular absorption~\cite{Sampei2023}. 
%% RESEARCH GAP
\bigskip\par
Although these results suggest the potential of Ising machines to be beneficial in certain engineering problems, especially in optimization tasks, there is still a need for further exploration of suitable use cases. Regarding the specific field of topology optimization, structural design has mainly been addressed, and, to the best of our knowledge, the application of Ising machine methods to the topology optimization of flow channels has not yet been investigated. 
\par
%% PURPOSE, OUTLINE, AND OBJECTIVES OF THIS WORK
To address this gap in the literature, the purpose of this study is to develop and test a design update strategy using a novel Ising machine formulation for the optimization of flow channels. 
Previously only addressed using classical approaches~\cite{Borrvall2003,Challis2009,Duan2008,Zhou2008}, we now present the first such formulation for topology optimization in fluid flow problems.
In particular, we introduce a \gls{qubo} formulation for the problem of finding a design update minimizing the energy dissipation of a flow channel subject to volume constraints for the material.
This update strategy is then integrated into the typical two-step optimization loop, where we first compute the state variables, in this case, the flow field, and then update the design in a second step. 
While the design update is often performed according to gradient information or heuristics, in this work, we base it on the solution of the \gls{qubo} problem, yielding intermediate optimal designs at each step.
\par
The specific objectives of this research are to evaluate the developed Ising machine formulation through numerical experiments:
\begin{enumerate} 
\item Analyze different terms in the formulation to understand their impact on the solution and the final design. 
\item Compare the results obtained using the Ising machine formulation with those from a traditional optimization method in terms of the number of optimization steps required and the solution quality of the resulting designs. 
\end{enumerate}
\bigskip\par
The remainder of this manuscript is organized as follows.
First, we provide a brief overview of the topology optimization for flow problems and explain the level-set-based material representation, which is followed by an outline of the problem of minimizing energy dissipation in flow channels. We then detail the proposed optimization approach, where the design update is performed using an Ising machine formulation.
Subsequently, we present two numerical test cases to investigate specific terms in the formulation and compare the performance of the approach with a classical method. Finally, we draw conclusions from the obtained results and provide an outlook, including future research directions.

%%%%%%%%%%%%%
%% METHODS %%
%%%%%%%%%%%%%
\section*{Methods}
\label{sec:methods}
In this section, we outline our approach for an update strategy in the topology optimization of flow channels using an Ising machine formulation. 
First, we provide a brief overview of topology optimization in the context of fluid flow problems.
Subsequently, we provide more details on the representation of the topology, i.e., the material distribution, using the level-set method before we state the problem of minimizing the dissipation energy in flow channels. 
Finally, we present the optimization approach in which the developed Ising machine formulation is integrated to compute design updates.

%% TOPOLOGY OPTIMIZATION FOR FLOW PROBLEMS
%% ---------------------------------------
\subsection*{Topology Optimization for Flow Problems}
Topology optimization aims to find the best design by optimizing the distribution of material in a given and fixed design domain $\domain$. In the context of flow problems, we consider regions of $\domain$ that contain material or void and refer to them as the solid region $\domainSolid$ or the fluid region $\domainFluid=\domain\backslash\domainSolid$, respectively.
To describe the distribution of the material in $\domain$, one can use a characteristic function that identifies $\domainSolid$, indicating the parts of $\domain$ where the material is located.
More specifically, given a position $\coord\in\domain$, the characteristic function $\charFunc[\solid]$ is represented as
\begin{equation}
\charFunc[\solid]\lp\coord\rp 
= 
\begin{cases}
    1, & \coord \in \domainSolid,\\
    0, & \coord \in \domainFluid.
\end{cases}
\end{equation}   
Using the characteristic function, we can, in a general setting, define an objective function that is maximized or minimized to obtain optimal structures.
We note that the formulation using the characteristic function enables flexible structural optimization. 
However, in structural topology optimization, the issue of discontinuous material distributions, such as checkerboard patterns, arises and needs to be addressed~\cite{Yamada2010}. 
%As a remedy for the issue, several approaches have been proposed, including the homogenization method, the density approach, and the level-set method~\cite{Bendsoe2013}. 
%As mentioned above, we will use the latter method and provide a detailed explanation in the following section.
\bigskip\par
To date, a number of attempts have been made to utilize topology optimization techniques for flow problems. 
For instance, it has been demonstrated that topology optimization can be applied to minimum power dissipation problems in Stokes flow~\cite{Borrvall2003} and steady and unsteady Navier-Stokes flows\cite{Olesen2006, Deng2011}, or drag minimization and lift maximization problems~\cite{Kondoh2012}. 
Moreover, its applicability to fluid devices~\cite{Gersborg2005}, thermal-fluid problems~\cite{Papoutsis2011}, and fluid-structure problems~\cite{Yoon2010} has been demonstrated.
\par
To account for the existence of material in fluid flow problems, the common approach is to consider the material to be a porous medium and add a corresponding resistance term to the governing equations. This resistance term is zero for fluid regions but non-zero for solid regions. In particular, the distinction between solid and fluid regions can be achieved by the local resistance coefficient $\resistanceCoeff\lp\coord\rp$, which is the inverse of the local permeability of the medium.
Then, we consider the generalized Stokes equations for the fluid velocity $\velocity$ and the pressure $\pressure$ including the resistance term $\resistanceCoeff\velocity$:
\begin{align}
    \nabla \cdot \velocity &= 0\quad\text{in }\Omega, 
    \label{eq:continuity}
    \\
    - \nabla \cdot \lp2\viscosity \strainRateTensor\lp\velocity\rp\rp + \alpha \velocity + \nabla \pressure &= 0\quad\text{in }\Omega,
    \label{eq:momentum}
\end{align}
where $\viscosity$ denotes the dynamic viscosity and $\strainRateTensor(\velocity) = (\nabla \velocity + (\nabla \velocity)^\mathsf{T})/2$ is the strain rate tensor.
We assume that the velocity $\velocity_D$ is prescribed on the Dirichlet portion of the domain boundary, $\domainBoundaryDirichlet$, and a homogeneous condition is applied on the Neumann portion of the domain boundary, $\domainBoundaryNeumann$:
\begin{align}
    \velocity &= \velocity_D \quad\text{on }\domainBoundaryDirichlet, 
    \label{eq:bcDirichlet}
    \\
    \lp-\pressure I +2\viscosity \strainRateTensor\lp\velocity\rp\rp \cdot  \normal &= 0\quad\text{on }\domainBoundaryNeumann ,
    \label{eq:bcNeumann}
\end{align}
where $\normal$ is the unit normal vector.
To determine the resistance coefficient, previous works have employed both the density approach and the level-set method. In the following, we will focus on the level-set method.

%% LEVEL-SET-BASED REPRESENTATION OF THE MATERIAL DISTRIBUTION
%% -----------------------------------------------------------
\subsection*{Level-Set-Based Representation of the Material Distribution}
As previously mentioned, a challenge in topology optimization is avoiding discontinuous material distributions that can arise from the introduction of the characteristic function $\charFunc[\solid]$. 
We explained that approaches exist to address this issue, including the homogenization method and the density approach~\cite{Bendsoe2013}. However, a potential downside of these methods is the so-called \textit{grayscale problem}, where intermediate fluid-solid regions are permitted in the optimal configuration~\cite{Yamada2010}. This can lead to ambiguous boundaries between solid and fluid regions, which may result in unrealistic structural designs.
\par
Another approach in topology optimization is the level-set method~\cite{Allaire2002,Wang2003,Allaire2004,Yamada2010}, which uses a level-set function to represent the interface between fluid and solid regions, thereby defining both regions. 
The scalar level-set function $\levelSet(\coord)\in[-1,1]$ is defined as follows:
\begin{equation}
    \begin{cases}
        -1 \le \levelSet\lp\coord\rp < 0    & \coord \in \domainSolid,\\
        \levelSet\lp\coord\rp = 0           & \coord \in \interface, \\
        0<\levelSet\lp\coord\rp\le1         & \coord \in \domainFluid,
    \end{cases}
\end{equation}   
where $\interface$ represents the interface between the fluid and solid regions.
In other words, the regions are determined by the sign of the level-set function, with the region where the function equals zero being regarded as the interface. 
Note that we can relate the level-set function $\levelSet$ to a characteristic function $\charFunc[\fluid]$ that indicates the fluid region such that
\begin{equation}
\charFunc[\fluid]\lp\coord\rp 
= 
\begin{cases}
    1, & \levelSet\lp\coord\rp > 0 \Leftrightarrow \coord \in \domainFluid,\\
    0, & \levelSet\lp\coord\rp < 0 \Leftrightarrow \coord \in \domainSolid.
\end{cases}
\label{eq:relationLevelSetCharFunc}
\end{equation}  
As the level-set function $\levelSet$, or equivalently the characteristic function $\charFunc[\fluid]$, locally distinguishes between the fluid and solid regions, we use $\charFunc[\fluid](\coord)$ to determine the properties of the porous medium, i.e., the local resistance coefficient $\resistanceCoeff(\coord)$.
More precisely, we define the resistance coefficient as
\begin{equation}
    \resistanceCoeff(\coord) = \resistanceCoeffSolid\lp1-\charFunc[\fluid]\lp\coord\rp\rp,
    \label{eq:resistanceCoeff}
\end{equation}
where $\resistanceCoeffSolid$ is the resistance coefficient for the solid region. 
Choosing a high value for $\resistanceCoeffSolid$ indicates that the region is nearly impermeable and can thus be considered solid. Therefore, $\resistanceCoeffSolid$ should be sufficiently large. Additionally, it is worth noting that the characteristic function is essentially the Heaviside function or, equivalently, a step function.

%% MINIMIZATION OF THE ENERGY DISSIPATION IN FLOW CHANNELS
%% -------------------------------------------------------
\subsection*{Minimization of the Energy Dissipation in Flow channels}
Next, we consider the scenario where the goal is to design a flow channel that minimizes energy dissipation, subject to a volume constraint for the fluid domain.
We approach this task using the level-set method explained in the previous section. 
In particular, we define a fixed domain $\Omega$ that is full of porous medium and differentiate between the fluid and solid regions according to $\charFunc[\fluid]$ and the resulting resistance coefficients $\resistanceCoeff(\charFunc[\fluid])$. 
\par
In this context, energy dissipation occurs due to two main factors: the viscosity of the fluid and the resistance from the porous medium. The viscosity of the fluid causes energy dissipation through internal friction, resulting in heat generation and a loss of mechanical energy. Additionally, the flow encounters resistance when moving through the porous medium, which is characterized by a high resistance coefficient, further contributing to energy dissipation.
Consequently, we can introduce the following objective function $\objective$ for the dissipation energy, which we aim to minimize during the optimization:
\begin{equation} 
    \objective (\charFunc[\fluid])
    = 
    \int_{\domain} 2 \viscosity \strainRateTensor\lp\velocity\rp:\strainRateTensor\lp\velocity\rp 
    + \resistanceCoeff(\charFunc[\fluid]) \velocity \cdot \velocity
    \,d\domain.
    \label{eq:objectiveFunctionDissipationEnergy}
\end{equation}
where $:$ is the operator that linearly maps a second-order tensor to a second-order tensor using a fourth-order tensor. In the case of an internal flow, this energy can also be regarded as the pressure drop between the inlet and outlet of the flow channel.
\par
Following Ref.~\cite{Yamada2010}, we use a regularized objective function $\objectiveRegularized = \objective + \regularization$ to avoid the discontinuity problem in topology optimization. To this end, the regularization term is given as
\begin{equation} 
    \regularization\lp\levelSet\rp= \frac{1}{2}\regularizationCoeff \int_{\domain} |\nabla \levelSet\lp\coord\rp|^2 \,d\domain,
    \label{eq:regularization}
\end{equation}
with corresponding coefficient $\regularizationCoeff$. 
\par
In addition to the objective of minimizing the term in \Cref{eq:objectiveFunctionDissipationEnergy}, we impose a constraint on the volume of the fluid region
\begin{equation}
    \volumeFluid = \int_{\domain} \charFunc[\fluid]\lp\coord\rp\,d\domain,
    \label{eq:volumeFluid}
\end{equation}
by requiring the size of the fluid region to take a constant target value $\volumeLimit \in \mathbb{R}$.
\par
Using these expressions, the optimization problem for the dissipation energy minimization task can be stated as follows:
\begin{subnumcases}{  \label{eq:optimizationProblem}}
    \inf\limits_{\charFunc[\fluid],\levelSet} \text{ } \objectiveRegularized(\charFunc[\fluid],\levelSet) = \inf\limits_{\charFunc[\fluid],\levelSet}\text{ }
    \lp
    \objective(\charFunc[\fluid]) + \regularization\lp\levelSet\rp
    \rp,
    \label{eq:objectiveRegularized}
    \\
    \text{s.t. }  \volumeFluid = \volumeLimit,
    \label{eq:volumeConstraint}
    \\
    \phantom{\text{s.t. } } A(\bm{u},p,\charFunc[\fluid]) =0, 
    \label{eq:momentumWeak}
    \\
    \phantom{\text{s.t. } }B(\bm{u}) = 0, 
    \label{eq:continuityWeak}
\end{subnumcases}
where $B(\velocity)$ and $A(\velocity,\pressure,\charFunc[\fluid])$ are the weak form representations of \Cref{eq:continuity,eq:momentum}, respectively, when applying the \gls{fem}. 

%% OPTIMIZATION APPROACH AND ISING MACHINE FORMULATION
%% ---------------------------------------------------
\subsection*{Optimization Approach and Ising Machine Formulation}
To solve the optimization problem given in \Cref{eq:optimizationProblem}, one can take a two-step procedure as employed in Refs.~\cite{Talischi2012,Pereira2016}: 
\begin{enumerate}
    \item Compute the flow field, i.e., the velocity $\velocity$ and the pressure $\pressure$, by the \gls{fem} according to the governing equations from \Cref{eq:continuity,eq:momentum,eq:bcDirichlet,eq:bcNeumann}. 
    \item With $\velocity$ and $\pressure$ fixed, update the level-set function $\levelSet$ and the characteristic function $\charFunc[\fluid]$ to reduce the objective function $\objectiveRegularized$ and improve the current design while respecting the volume constraint. 
\end{enumerate}
These two steps are executed iteratively until a terminating condition is met, such as when the change in the objective function in a certain iteration becomes negligibly small.
Note that each iteration requires one evaluation of the \gls{fem} to compute the flow field for the latest design.
\par
One option to update the design is to follow a gradient-based approach, utilizing information on how the objective function changes with design modifications. For details on this optimization process, please refer to Refs.~\cite{Talischi2012, Pereira2016}.
We note that one can also introduce the method of Lagrange multipliers to solve the optimization problem, where the governing equations and a time evolution problem for the level-set function are iteratively solved using the sensitivity of the objective function~\cite{Yamada2010,Yaji2014}.
\par
In this work, however, we follow an alternative approach.
While we also use a two-step procedure as explained above and leave the first step, i.e., the computation of the flow field by the \gls{fem}, unchanged, we focus on how to compute the design update in the second step. In particular, based on the given quantities $\velocity$ and $\pressure$ from the first step, we solve a problem in an Ising machine formulation to update the design, i.e., the level-set function $\levelSet$ and the characteristic function $\charFunc[\fluid]$.
As mentioned before, the motivation is to investigate if the overall optimization procedure can be accelerated by innovative optimization techniques such as Ising machines.
\par
For the Ising machine formulation, we derive a \gls{qubo} problem corresponding to the minimization of Equation \eqref{eq:objectiveRegularized} while considering the volume constraint from Equation \eqref{eq:volumeConstraint}.
Note that Equation \eqref{eq:objectiveRegularized} involves the energy dissipation term from Equation \eqref{eq:objectiveFunctionDissipationEnergy} and the regularization term from Equation \eqref{eq:regularization}. All of the aforementioned terms are, for fixed $\velocity$ and $\pressure$, only a function of the level-set field $\levelSet$ or the characteristic function $\charFunc[\fluid]$.
Therefore, we derive an objective function in \gls{qubo} form that accounts for each of these terms and only depends on $\levelSet$ and $\charFunc[\fluid]$.
\bigskip\par
In the following, we will assume that we have a subdivision of the domain into elements $\elemIndex$. Then, we consider the element-wise level-set and characteristic functions denoted by $\levelSetElem$ and $\charFuncElem[\fluid]$, respectively.
Since a \gls{qubo} problem must involve only binary variables, we have to express both functions entirely in terms of binary variables. Note that the characteristic function $\charFuncElem[\fluid]$ can be directly represented by a single binary variable. However, we need to find some representation for the real-valued level-set function $\levelSetElem$.
A straightforward approach is to use a binary system of notation, where any number $a\in\mathbb{R}$ is written as $a=\sum_{i=1}^{N} 2^{i_{0}-i} q_i$ using $N$ binary variables $q_i$ and a fixed integer $i_{0}$. 
Note that the way of expressing a real-valued variable through binary variables has been reported to influence the optimization performance of annealing machines~\cite{Endo2024}.
In this study, the so-called \textit{uniform-weighted sum representation} is used to express the variables for the level-set function $\levelSetElem\in[-1,1]$ trough corresponding binary variables $q^k_i$:  
\begin{equation}
    \levelSetElem\lp q^k_i\rp
    = 
    2 \left( \frac{1}{N} \sum_{i=1}^{N} q^{k}_{i}\right) -1. 
\end{equation}
Moreover, we use one additional binary variable per element $q^{k}_{\chi}$ for the characteristic function
\begin{equation}
    \charFuncElem[\fluid]\lp q^{k}_{\chi} \rp = q^{k}_{\chi},
\end{equation}
which is used for computing the resistance coefficient $\resistanceCoeff$ from \Cref{eq:resistanceCoeff} and the volume of the fluid region $\volumeFluid$ from \Cref{eq:volumeFluid}. 
Consequently, each element $k$ holds $N+1$ binary variables $\{q^k_1, q^k_2,\dots,q^k_N,q^k_{\chi}\}$.
Using this, we define the objective function $\objectiveQUBO(\charFuncElem[\fluid],\levelSetElem)$ for the \gls{qubo} problem as follows:
\begin{equation} 
    \objectiveQUBO\lp\charFuncElem[\fluid],\levelSetElem\rp 
    = 
    \objectiveDissipation\lp\charFuncElem[\fluid]\rp  
    + 
    \objectiveRegularization\lp\levelSetElem\rp 
    + 
    \objectiveVolume\lp\charFuncElem[\fluid]\rp  
    + 
    \objectiveCharacteristic\lp\charFuncElem[\fluid],\levelSetElem\rp ,
    \label{eq:objectiveQUBO}
\end{equation}
with contributions corresponding to the energy dissipation term in \Cref{eq:objectiveFunctionDissipationEnergy}
\begin{align}
    \objectiveDissipation\lp\charFuncElem[\fluid]\rp 
    &=
    \paramDissipation \int_{\domain} 
    %\strainRateTensor\lp\velocity\rp:\strainRateTensor\lp\velocity\rp + 
    \resistanceCoeff\lp\charFuncElem[\fluid]\rp \velocity \cdot \velocity\,d\domain,
    \label{eq:qubo_dis}
    \intertext{the regularization term in \Cref{eq:regularization}}
    \objectiveRegularization\lp\levelSetElem\rp 
    &= 
    \paramRegularization \int_{\domain}|\nabla\levelSetElem\lp\coord\rp|^2\,d\domain, 
    \label{eq:qubo_reg}
    \intertext{the volume constraint from Equation \eqref{eq:volumeConstraint}}
    \objectiveVolume\lp\charFuncElem[\fluid]\rp 
    &= 
    \paramVolume\left( \volumeFluid - \volumeLimit\right)^2, 
    \label{eq:qubo_vol}
    \intertext{and a term that couples the level-set and the characteristic function}
    \objectiveCharacteristic\lp\charFuncElem[\fluid],\levelSetElem\rp 
    &= 
    \paramCharacteristic \int_{\domain} \left( \charFuncElem[\fluid]\lp\coord\rp- \frac{1}{2}\lp1+\levelSetElem\lp\coord\rp\right) \rp^2\,d\domain, 
    \label{eq:qubo_char}
\end{align}
with positive and real-valued hyperparameters $\paramDissipation,\paramRegularization,\paramVolume,\paramCharacteristic\in\mathbb{R}^{+}$.
\par
Note that in the energy dissipation term in \Cref{eq:qubo_dis}, we can omit the viscous contribution, i.e., the first term in \Cref{eq:objectiveFunctionDissipationEnergy}, as it only adds a constant for a given velocity $\velocity$.
Furthermore, we can absorb prefactors and coefficients, such as the coefficient $\regularizationCoeff$ appearing in the original regularization term in \Cref{eq:regularization}, into the corresponding hyperparameters.
Finally, the fourth term $\objectiveCharacteristic(\charFuncElem[\fluid],\levelSetElem)$ given in \Cref{eq:qubo_char} is introduced to ensure that the level-set and the characteristic functions are consistent, meaning they both indicate the same fluid region. To this end, we use the mean squared error between the characteristic function $\charFuncElem[\fluid]\in\{0,1\}$ and the rescaled level-set function $\frac{1}{2}(1+\levelSetElem(\coord))\in[0,1]$ to realize a penalization of any inconsistency.
\par
We recall that we take the two-step process and, thus, the velocity $\velocity$ and the pressure $\pressure$ are at hand when computing $\objectiveQUBO(\charFuncElem[\fluid],\levelSetElem)$.
In addition, we stress that these terms can be expressed as a quadratic form with respect to the binary variables used for the level-set function $\levelSetElem$ and the characteristic function $\charFuncElem[\fluid]$.
Therefore, the objective function $\objectiveQUBO$ indeed provides a \gls{qubo} problem, and we can perform the design update based on solving it on Ising machines. In particular, we obtain the updated design variables as
\begin{equation}
    \charFuncElemNew[\fluid],\levelSetElemNew
    =
    \argmin_{\charFuncElem[\fluid],\levelSetElem} \objectiveQUBO\lp\charFuncElem[\fluid],\levelSetElem\rp,
\end{equation}
or, equivalently, by solving the \gls{qubo} problem for the updated binary variables $\hat{q}^k_i$ and $\hat{q}^k_{\chi}$:
\begin{equation}
   \hat{q}^k_i,\hat{q}^k_{\chi} = \argmin_{q^k_i,q^k_{\chi}} \objectiveQUBO\lp\charFuncElem[\fluid]\lp q^k_{\chi}\rp,\levelSetElem\lp q^k_i\rp\rp.
\end{equation}
%%%%%%%%%%%%%%%%%%%%%%%%%%%%
%% RESULTS AND DISCUSSION %%
%%%%%%%%%%%%%%%%%%%%%%%%%%%%
\section*{Results and Discussion}
\label{sec:results}
\begin{figure}
    \centering
    % \resizebox{0.5\textwidth}{!}{
        \begin{tikzpicture}[scale=5]

    % DIFFUSER PROBLEM
    \begin{scope}
        % Draw the main square
        \draw (0,0) rectangle (1,1);

        \draw[-latex] (0,0) -- (0.15,0) node [below] {$x$};
        \draw[-latex] (0,0) -- (0,0.15) node [left] {$y$};

        % Add dimensions
        \draw[latex-latex] (0.9,0) -- (0.9,1/3) node[midway,left] {$\frac{1}{3}$};
        \draw[latex-latex] (0.9,1/3) -- (0.9,2/3) node[midway,left] {$\frac{1}{3}$};
        \draw[latex-latex] (0.9,2/3) -- (0.9,1) node[midway,left] {$\frac{1}{3}$};
        
        \draw[latex-latex] (0,-0.1) -- (1,-0.1) node[midway,below] {$1$};
        \draw[latex-latex] (-0.1,0) -- (-0.1,1) node[midway,left] {$1$};

        \draw[thick,domain=0:1] plot (-0.5*\x*\x+0.5*\x,\x);
        \foreach \y in {0.2, 0.4, ..., 0.8} {
            \draw[-latex] (0,\y) -- ({-0.5*\y*\y+0.5*\y},\y);
        }
        
        \draw[thick,domain=0.3333333:0.66666666] plot (1-13.5*\x*\x+13.5*\x-13.5*0.22222,\x);
        \foreach \y in {6/15, 7/15, 8/15, 9/15} {
            \draw[-latex] (1,\y) -- ({1-13.5*\y*\y+13.5*\y-13.5*0.22222},\y);
        }  
    \end{scope}

    % DOUBLE PIPE PROBLEM
    \begin{scope}[shift={(1.75,0)}]
        % Draw the main square
        \draw (0,0) rectangle (1.5,1);

        \draw[-latex] (0,0) -- (0.15,0) node [below] {$x$};
        \draw[-latex] (0,0) -- (0,0.15) node [left] {$y$};
        
        % Add dimensions
        \draw[latex-latex] (1.4,0) -- (1.4,1/6) node[midway,left] {$\frac{1}{6}$};
        \draw[latex-latex] (1.4,1/6) -- (1.4,2/6) node[midway,left] {$\frac{1}{6}$};
        \draw[latex-latex] (1.4,2/6) -- (1.4,4/6) node[midway,left] {$\frac{1}{3}$};
        \draw[latex-latex] (1.4,4/6) -- (1.4,5/6) node[midway,left] {$\frac{1}{6}$};
        \draw[latex-latex] (1.4,5/6) -- (1.4,1) node[midway,left] {$\frac{1}{6}$};
        
        \draw[latex-latex] (0,-0.1) -- (1.5,-0.1) node[midway,below] {$1.5$};
        \draw[latex-latex] (-0.1,0) -- (-0.1,1) node[midway,left] {$1$};

        \draw[thick,domain=1/6:2/6] plot (-18*\x*\x+18*0.5*\x-18*1/18,\x);
        \foreach \y in {6/30, 7/30, 8/30, 9/30} {
            \draw[-latex] (0,\y) -- ({-18*\y*\y+18*0.5*\y-18*1/18},\y);
        }

        \draw[thick,domain=4/6:5/6] plot (-18*\x*\x+18*9/6*\x-18*20/36,\x);
        \foreach \y in {21/30, 22/30, 23/30, 24/30} {
            \draw[-latex] (0,\y) -- ({-18*\y*\y+18*9/6*\y-18*20/36},\y);
        }

        \draw[thick,domain=1/6:2/6] plot (1.5-18*\x*\x+18*0.5*\x-18*1/18,\x);
        \foreach \y in {6/30, 7/30, 8/30, 9/30} {
            \draw[-latex] (1.5,\y) -- ({1.5-18*\y*\y+18*0.5*\y-18*1/18},\y);
        }

        \draw[thick,domain=4/6:5/6] plot (1.5-18*\x*\x+18*9/6*\x-18*20/36,\x);
        \foreach \y in {21/30, 22/30, 23/30, 24/30} {
            \draw[-latex] (1.5,\y) -- ({1.5-18*\y*\y+18*9/6*\y-18*20/36},\y);
        }
    \end{scope}

\end{tikzpicture}
    % }
    \caption{The problem setting of the diffuser problem (left) and the double pipe problem (right).}
    \label{fig:problemsetting}
\end{figure}
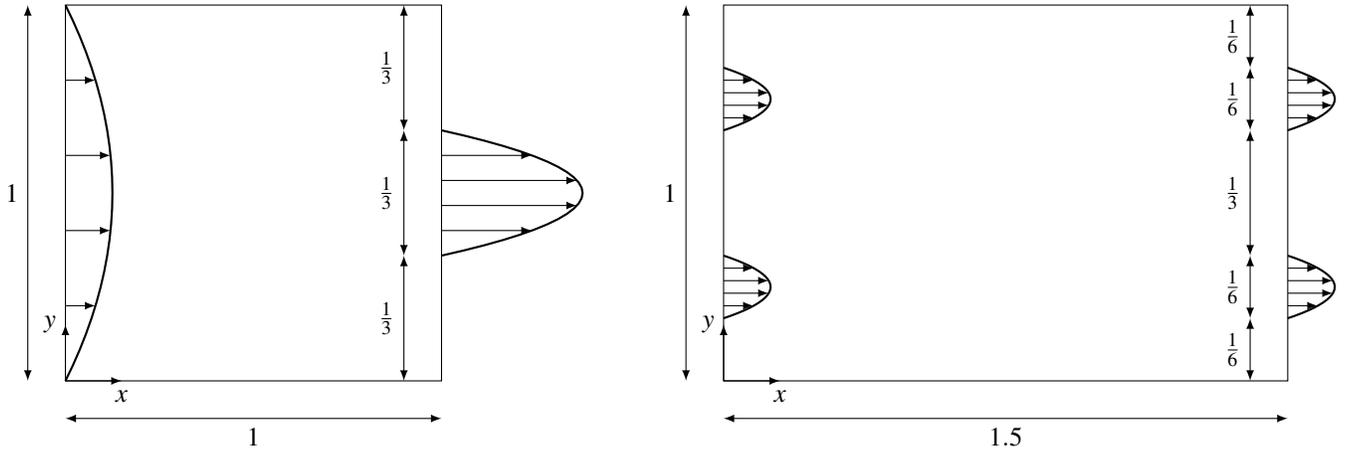

In the following, we present the results of numerical experiments conducted to test the optimization approach using the proposed update strategy based on the Ising machine formulation.
To this end, we consider two benchmark test cases commonly used in topology optimization for flow channels~\cite{Pereira2016,Yaji2014}: (1) the \textit{diffuser} problem and (2) the \textit{double pipe} problem.
The setup for the diffuser problem is shown in \Cref{fig:problemsetting} (left) and involves a square domain with an inlet along the left boundary and a smaller outlet on the right. 
In the case of the double pipe problem (see \Cref{fig:problemsetting} (right)), we consider a rectangular domain with two inlets and two outlets on the left and right boundaries, respectively.
For the discretization of the domain $\domain$, we use structured grids with quadrilateral elements. For the diffuser problem, the size of the mesh is 32 $\times$ 32 elements, while the mesh for the double pipe problem has 32 $\times$ 48 elements.
\par
To solve the \gls{qubo} problems in our approach, we use the \textit{Fixstars Amplify} \gls{ae}, a GPU-based Ising machine~\cite{Fixstars}. 
For the flow field computation using \gls{fem} and for comparison with the classical optimization approach, we integrated the method introduced in Ref.~\cite{Pereira2016} in our own implementation.
The code and the scripts used for the numerical experiments are available online~\cite{Key2024TopoFlow}.
\par
Using these test cases, we analyze the proposed optimization approach from two perspectives: (1) the study of individual contributions to the \gls{qubo} objective function $\objectiveQUBO$ and (2) a comparison with the classical method in terms of the number of optimization steps and the final objective function values for the dissipated energy $\objective$.
We will elaborate on the corresponding results in the following sections.
%% STUDY OF THE OBJECTIVE FUNCTION IN THE QUBO PROBLEM
%% -------------------------------------------------
\subsection*{Study of the Objective Function in the \acrshort{qubo} Problem}
The objective function in the \gls{qubo} problem $\objectiveQUBO$, given in \Cref{eq:objectiveQUBO}, has four different contributions: $\objectiveDissipation$, $\objectiveRegularization$, $\objectiveVolume$, and $\objectiveCharacteristic$. Consequently, the weighting of these contributions relative to each other through the corresponding hyperparameters will influence the final outcome of solving the \gls{qubo} problem. To gain a better understanding of the effects attributed to individual terms, we present the following results.
\par
\begin{figure}
    \centering
    % \resizebox{\textwidth}{!}{
        % This file was created with tikzplotlib v0.10.1.
\begin{tikzpicture}

\definecolor{darkgray176}{RGB}{176,176,176}
\definecolor{steelblue31119180}{RGB}{31,119,180}

\begin{axis}[
scale only axis,
height=150pt, % Increase height for y-axis
width=250pt,   % Keep width the same
tick align=outside,
tick pos=left,
title={Effect of \(\displaystyle \lambda_{\text{char}}\)},
x grid style={darkgray176},
xlabel={\(\displaystyle \lambda_{\text{char}}\)},
xmin=0.275, xmax=5.225,
xtick style={color=black},
y grid style={darkgray176},
ylabel={\# Inconsistent Elements},
ymin=-1, ymax=21,
ytick style={color=black}
]
\addplot [gray, dashed, thick] coordinates {(0.275, 0) (5.225, 0)};
\addplot [thick, steelblue31119180, mark=x, mark size=3, mark options={solid}]
table {%
0.5 20
1 10
1.5 5
2 2
2.5 2
3 0
3.5 0
4 0
4.5 0
5 0
};
\end{axis}

\end{tikzpicture}
    % }
    \caption{Diffuser problem: the number of elements $\elemIndex$ with inconsistent values of $\levelSetElem$ and $\charFuncElem[\fluid]$ for different values of $\paramCharacteristic$.}
    \label{fig:inconsistenciesDiffuser}
\end{figure}
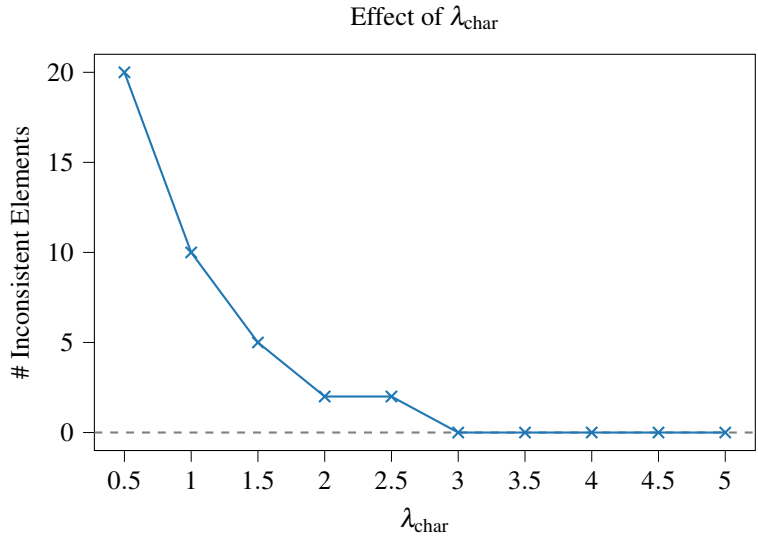

First, we focus on the term $\objectiveCharacteristic$ and the corresponding hyperparameter $\paramCharacteristic$, which serve to ensure the consistency between the level-set function $\levelSetElem$ and the characteristic function $\charFuncElem[\fluid]$ according to \Cref{eq:relationLevelSetCharFunc}. To this end, we vary $\paramCharacteristic$ and track the resulting number of elements on which inconsistent values of $\levelSetElem$ and $\charFuncElem[\fluid]$ occur.
We consider the diffuser problem and let $\paramCharacteristic$ run from $0.5$ to $5.0$, applying a step size of $0.5$. All the other hyperparameters are kept fixed as $\paramDissipation=100$, $\paramRegularization=1$, and $\paramVolume=20$.
In addition, we use $\viscosity=1$, $\resistanceCoeffSolid=12.5$, and $N=8$.
For the \gls{ae}, we set the time-out time $\timeOut$, i.e., the maximum time available for the solution of the \gls{qubo} problem, as $\timeOut = 1,000\,ms$.
The resulting graph is given in \Cref{fig:inconsistenciesDiffuser} and shows the number of inconsistencies for the different values of $\paramCharacteristic$. 
The behavior is as expected as the number of inconsistent elements decreases with increasing $\paramCharacteristic$ until it reaches the zero line.
\par
\begin{figure}
    \centering
    \begin{tikzpicture}
    \begin{groupplot}[group style={group size=5 by 2, horizontal sep=10pt, vertical sep=10pt},height=120pt,width=120pt]
        \nextgroupplot[
        scaled x ticks=manual:{}{\pgfmathparse{#1}},
        scaled y ticks=manual:{}{\pgfmathparse{#1}},
        tick align=outside,
        x grid style={darkgray176},
        xmajorticks=false,
        xmin=-0., xmax=1.,
        xtick style={color=black},
        xticklabels={},
        y dir=reverse,
        y grid style={darkgray176},
        ylabel style={rotate=-90.0},
        ylabel={$\levelSetElem$},
        ymajorticks=false,
        ymin=-0., ymax=1.,
        ytick style={color=black},
        yticklabels={}
        ]
        \addplot graphics [includegraphics cmd=\pgfimage,xmin=0, xmax=1, ymin=0, ymax=1] {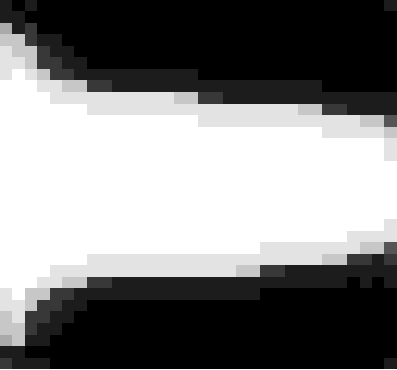};
        
        \nextgroupplot[
        scaled x ticks=manual:{}{\pgfmathparse{#1}},
        scaled y ticks=manual:{}{\pgfmathparse{#1}},
        tick align=outside,
        x grid style={darkgray176},
        xmajorticks=false,
        xmin=-0., xmax=1.,
        xtick style={color=black},
        xticklabels={},
        y dir=reverse,
        y grid style={darkgray176},
        ylabel style={rotate=-90.0},
        ylabel={},
        ymajorticks=false,
        ymin=-0., ymax=1.,
        ytick style={color=black},
        yticklabels={}
        ]
        \addplot graphics [includegraphics cmd=\pgfimage,xmin=0, xmax=1, ymin=0, ymax=1] {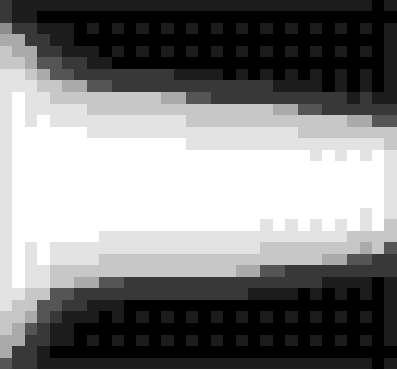};

        \nextgroupplot[
        scaled x ticks=manual:{}{\pgfmathparse{#1}},
        scaled y ticks=manual:{}{\pgfmathparse{#1}},
        tick align=outside,
        x grid style={darkgray176},
        xmajorticks=false,
        xmin=-0., xmax=1.,
        xtick style={color=black},
        xticklabels={},
        y dir=reverse,
        y grid style={darkgray176},
        ylabel style={rotate=-90.0},
        ylabel={},
        ymajorticks=false,
        ymin=-0., ymax=1.,
        ytick style={color=black},
        yticklabels={}
        ]
        \addplot graphics [includegraphics cmd=\pgfimage,xmin=0, xmax=1, ymin=0, ymax=1] {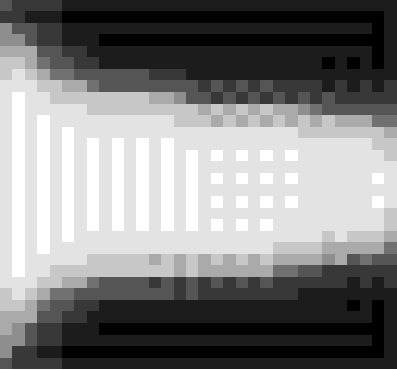};

        \nextgroupplot[
        scaled x ticks=manual:{}{\pgfmathparse{#1}},
        scaled y ticks=manual:{}{\pgfmathparse{#1}},
        tick align=outside,
        x grid style={darkgray176},
        xmajorticks=false,
        xmin=-0., xmax=1.,
        xtick style={color=black},
        xticklabels={},
        y dir=reverse,
        y grid style={darkgray176},
        ylabel style={rotate=-90.0},
        ylabel={},
        ymajorticks=false,
        ymin=-0., ymax=1.,
        ytick style={color=black},
        yticklabels={}
        ]
        \addplot graphics [includegraphics cmd=\pgfimage,xmin=0, xmax=1, ymin=0, ymax=1] {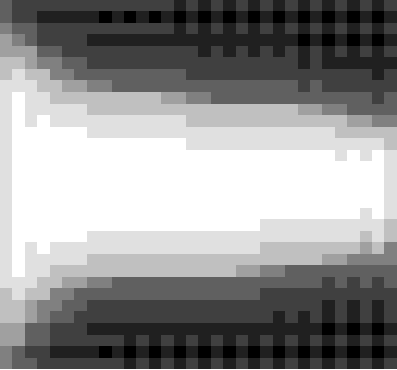};

        \nextgroupplot[
        colorbar,
        colorbar style={ylabel={}},
        colormap/blackwhite,
        point meta min=0,
        point meta max=1,
        scaled x ticks=manual:{}{\pgfmathparse{#1}},
        scaled y ticks=manual:{}{\pgfmathparse{#1}},
        tick align=outside,
        x grid style={darkgray176},
        xmajorticks=false,
        xmin=-0., xmax=1.,
        xtick style={color=black},
        xticklabels={},
        y dir=reverse,
        y grid style={darkgray176},
        ylabel style={rotate=-90.0},
        ylabel={},
        ymajorticks=false,
        ymin=-0., ymax=1.,
        ytick style={color=black},
        yticklabels={}
        ]
        \addplot graphics [includegraphics cmd=\pgfimage,xmin=0, xmax=1, ymin=0, ymax=1] {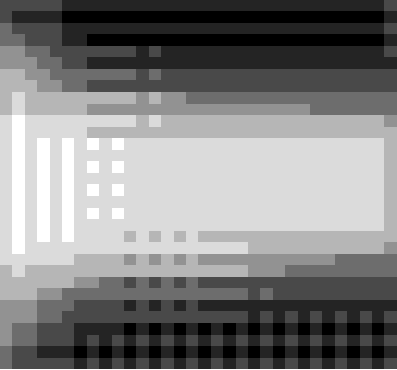};        

        \nextgroupplot[
        scaled x ticks=manual:{}{\pgfmathparse{#1}},
        scaled y ticks=manual:{}{\pgfmathparse{#1}},
        tick align=outside,
        x grid style={darkgray176},
        xmajorticks=false,
        xmin=-0., xmax=1.,
        xtick style={color=black},
        xticklabels={},
        y dir=reverse,
        y grid style={darkgray176},
        ylabel style={rotate=-90.0},
        ylabel={$\charFuncElem[\levelSet]$},
        ymajorticks=false,
        ymin=-0., ymax=1.,
        ytick style={color=black},
        yticklabels={}
        ]
        \addplot graphics [includegraphics cmd=\pgfimage,xmin=0, xmax=1, ymin=0, ymax=1] {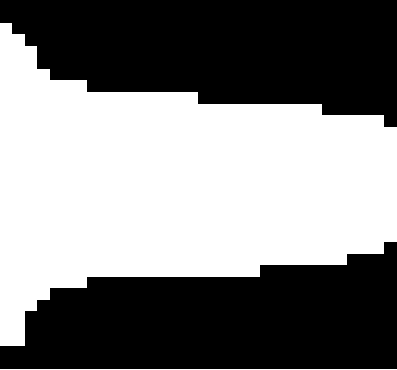};
        
        \nextgroupplot[
        scaled x ticks=manual:{}{\pgfmathparse{#1}},
        scaled y ticks=manual:{}{\pgfmathparse{#1}},
        tick align=outside,
        x grid style={darkgray176},
        xmajorticks=false,
        xmin=-0., xmax=1.,
        xtick style={color=black},
        xticklabels={},
        y dir=reverse,
        y grid style={darkgray176},
        ylabel style={rotate=-90.0},
        ylabel={},
        ymajorticks=false,
        ymin=-0., ymax=1.,
        ytick style={color=black},
        yticklabels={}
        ]
        \addplot graphics [includegraphics cmd=\pgfimage,xmin=0, xmax=1, ymin=0, ymax=1] {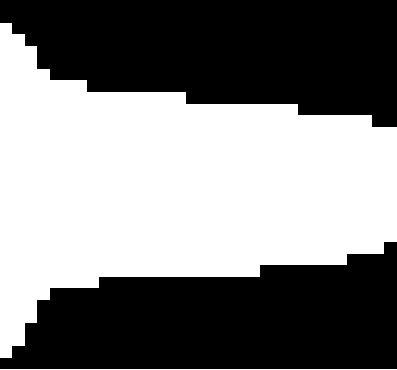};

        \nextgroupplot[
        scaled x ticks=manual:{}{\pgfmathparse{#1}},
        scaled y ticks=manual:{}{\pgfmathparse{#1}},
        tick align=outside,
        x grid style={darkgray176},
        xmajorticks=false,
        xmin=-0., xmax=1.,
        xtick style={color=black},
        xticklabels={},
        xlabel ={$\paramRegularization\longrightarrow$},
        y dir=reverse,
        y grid style={darkgray176},
        ylabel style={rotate=-90.0},
        ylabel={},
        ymajorticks=false,
        ymin=-0., ymax=1.,
        ytick style={color=black},
        yticklabels={}
        ]
        \addplot graphics [includegraphics cmd=\pgfimage,xmin=0, xmax=1, ymin=0, ymax=1] {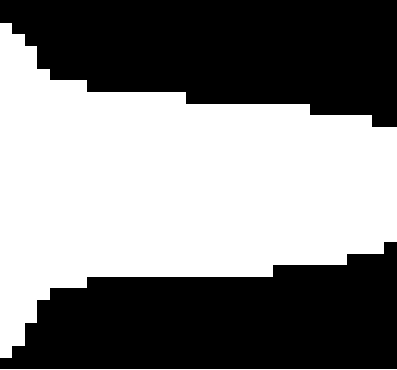};

        \nextgroupplot[
        scaled x ticks=manual:{}{\pgfmathparse{#1}},
        scaled y ticks=manual:{}{\pgfmathparse{#1}},
        tick align=outside,
        x grid style={darkgray176},
        xmajorticks=false,
        xmin=-0., xmax=1.,
        xtick style={color=black},
        xticklabels={},
        y dir=reverse,
        y grid style={darkgray176},
        ylabel style={rotate=-90.0},
        ylabel={},
        ymajorticks=false,
        ymin=-0., ymax=1.,
        ytick style={color=black},
        yticklabels={}
        ]
        \addplot graphics [includegraphics cmd=\pgfimage,xmin=0, xmax=1, ymin=0, ymax=1] {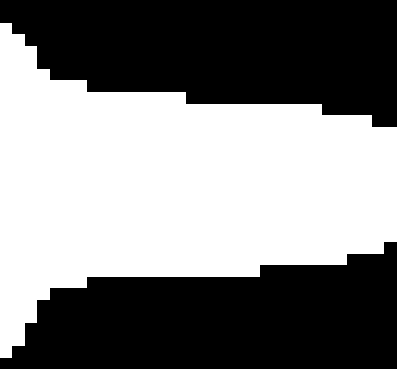};

        \nextgroupplot[
        colorbar,
        colorbar style={ylabel={}},
        colormap/blackwhite,
        point meta min=0,
        point meta max=1,
        scaled x ticks=manual:{}{\pgfmathparse{#1}},
        scaled y ticks=manual:{}{\pgfmathparse{#1}},
        tick align=outside,
        x grid style={darkgray176},
        xmajorticks=false,
        xmin=-0., xmax=1.,
        xtick style={color=black},
        xticklabels={},
        y dir=reverse,
        y grid style={darkgray176},
        ylabel style={rotate=-90.0},
        ylabel={},
        ymajorticks=false,
        ymin=-0., ymax=1.,
        ytick style={color=black},
        yticklabels={}
        ]
        \addplot graphics [includegraphics cmd=\pgfimage,xmin=0, xmax=1, ymin=0, ymax=1] {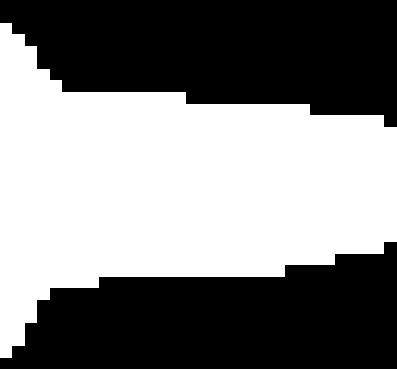};
    \end{groupplot}
\end{tikzpicture}
    \caption{Diffuser problem: effect of the regularization term on the level-set function $\levelSetElem$ (top) and the characteristic function $\charFuncElem[\fluid]$ (bottom) through different values of $\paramRegularization$, increasing from left to right.}
    \label{fig:diffuserEffectRegularization}
\end{figure}
Next, we consider the effect of the regularization term $\objectiveRegularization$ with its weight $\paramRegularization$.
Again, we use the diffuser problem and track the influence of varying $\paramRegularization$ on the final result.
In \Cref{fig:diffuserEffectRegularization}, the final distributions for the level-set function $\levelSetElem$ and the characteristic function $\charFuncElem[\fluid]$ are depicted for different values of $\paramRegularization$. 
On the one hand, we observe that increasing the weight for the regularization term leads to smoother distributions of the level-set field (top), which is the intended effect. On the other hand, the regularization does not change the final distribution of the characteristic function (bottom) significantly. This raises the question of the necessity of the regularization term in the current context. 
\par
Guided by this question, we further investigate the impact of the regularization term $\objectiveRegularization$ on the final design. In particular, we compare it to that of the original objective $\objectiveDissipation$, i.e., the energy dissipation term. To this end, we perform a parameter study including the weights for the regularization $\paramRegularization$ and the energy dissipation $\paramDissipation$.
The final designs for each parameter sample are shown in \Cref{fig:disVsReg}. As suggested by the previous results, we note that different weights for the regularization term $\paramRegularization$ (fixed row) do not lead to significantly different final designs. In contrast, varying $\paramDissipation$ (fixed column) clearly influences the character of the resulting material distribution. Specifically, increasing $\paramDissipation$ yields smoother designs without fluid inclusions in the solid. 
\begin{figure}
    \centering
    \input{fig/parameter_study/diffuser/regularizationDissipation/dis_reg}
    \caption{Diffuser problem: final designs for different values of $\paramDissipation$ (rows) and $\paramRegularization$ (columns).}
    \label{fig:disVsReg}
\end{figure}
\bigskip\par
Based on these results, we conclude that the regularization term is not necessary for the topology optimization of flow channels but the objective of minimizing energy dissipation already leads to continuous material distribution, which is consistent with existing literature~\cite{Pereira2016}. This is in contrast to the topology optimization in structural mechanics for which the regularization term has been originally introduced~\cite{Yamada2010}.
That can also be understood through an illustrative explanation.
In structural mechanics, voids in the structure, i.e., the absence of material, can be favorable for reducing weight. If those regions do not contribute to the load-bearing of the structure, its mechanical performance remains unaffected. However, when dealing with flow channels, such voids are not favorable since the continuity of the channel is crucial for fluid transport. Any voids would mean that fluid cannot pass those regions, leading to stagnation zones.
\par
This result has the following consequences for our approach. Since it is only the regularization term $\objectiveRegularization(\levelSetElem)$ for which we need the level-set field $\levelSetElem$, we can drop $\levelSetElem$ and, additionally, the term $\objectiveCharacteristic$, which had the purpose to ensure consistency between the level-set field $\levelSetElem$ and the characteristic function $\charFuncElem[\fluid]$. So, in the following, we will consider a condensed \gls{qubo} problem with the objective function
\begin{equation}
    \objectiveQUBOCondensed\lp\charFuncElem[\fluid]\rp
    =
    \objectiveDissipation\lp\charFuncElem[\fluid]\rp
    +
    \paramVolume\objectiveVolume\lp\charFuncElem[\fluid]\rp.
\end{equation}
Consequently, we only need one single binary variable per element in our formulation since all the binary variables previously related to the real-valued level-set field are no longer needed. This means that the number of required binary variables can be drastically decreased, being favorable for the efficiency of the approach.
Using the condensed formulation, we will compare the performance of the approach with a classical method in the next section.
%% COMPARISON WITH A CLASSICAL OPTIMIZATION APPROACH
%% -------------------------------------------------
\subsection*{Comparison with a Classical Optimization Approach}
To test the performance of the proposed approach, we use the two benchmark test cases introduced above to compare the results with those from a classical optimization approach as a reference. 
Here, our focus is on both the computational time and the quality of the optimized design, i.e., the number of optimization steps and the value of the objective function, respectively.
There are two main reasons why we chose the number of optimization steps as a measure of computational time.
First, the overall computational time for the optimization is often dominated by the time needed to solve the state equations, e.g., by the \gls{fem}, which scales with the number of optimization steps. 
Furthermore, the total computational time can be expressed as the product of the number of optimization steps and the time taken for each step. Since the latter heavily depends on the efficiency of the implementation and the hardware resources in use, it is not suitable for comparison.
\par
As previously mentioned, we pick the classical approach introduced in Ref.~\cite{Pereira2016}, which is based on the density method to represent the material distribution.
This approach also follows the two-step optimization process to minimize energy dissipation, allowing for a comparison in terms of the number of optimization steps as well as the values of the objective function.
To enable a fair comparison, we use the relative change in the value of the objective function $\objective$ as the termination criterion for the optimization in both cases. Furthermore, for the classical approach, we filter the final distribution of the real-valued design variable $\rho\in[0,1]$ used to obtain a pure binary representation. In particular, we consider every element in which $\rho>0.95$ holds as void.
\bigskip\par
First, we focus on the results of the diffuser test case. Here, we set the maximum volume to $\volumeLimit=\frac{1}{2}|\domain|$, where $|\domain|$ denotes the volume of the entire design domain.
As the initial design, we use no material at all, i.e., $\domain=\domainFluid$. The values for the material properties are $\viscosity=1.0$ and $\resistanceCoeffSolid=12.5$, while the hyperparameter from the condensed \gls{qubo} problem is given as $\paramVolume=0.2$.
As before, we set the time-out period to $\timeOut = 1,000\,ms$.
\par
\begin{figure}
    \centering
    \resizebox{0.45\textwidth}{!}{
        % This file was created with tikzplotlib v0.10.1.
\begin{tikzpicture}

\definecolor{darkgray176}{RGB}{176,176,176}

\begin{axis}[
tick align=outside,
tick pos=left,
title={Classical},
x grid style={darkgray176},
xlabel={$x$},
xmin=0, xmax=1,
xtick style={color=black},
y grid style={darkgray176},
ylabel={$y$},
ymin=0, ymax=1,
ytick style={color=black}
]
\addplot graphics [includegraphics cmd=\pgfimage,xmin=0, xmax=1, ymin=0, ymax=1] {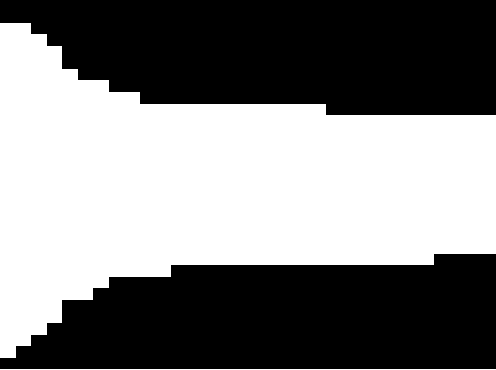};
\end{axis}

\end{tikzpicture}
    }
    \hfill
    \resizebox{0.45\textwidth}{!}{
        % This file was created with tikzplotlib v0.10.1.
\begin{tikzpicture}

\definecolor{darkgray176}{RGB}{176,176,176}

\begin{axis}[
tick align=outside,
tick pos=left,
title={Annealing},
x grid style={darkgray176},
xlabel={$x$},
xmin=0, xmax=1,
xtick style={color=black},
y grid style={darkgray176},
ylabel={$y$},
ymin=0, ymax=1,
ytick style={color=black}
]
\addplot graphics [includegraphics cmd=\pgfimage,xmin=0, xmax=1, ymin=0, ymax=1] {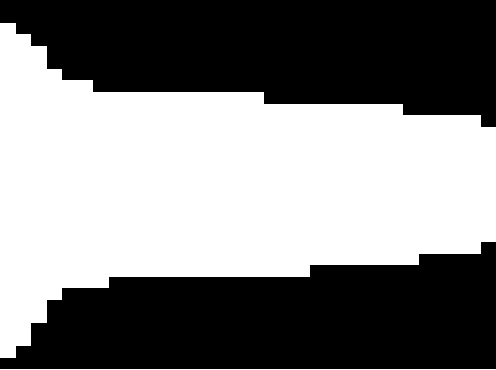};
\end{axis}

\end{tikzpicture}
    }
    \caption{Diffuser problem: final designs for the classical approach (left) and the annealing approach (right).
    }
    \label{fig:diffuser_final_designs}
\end{figure}

In \Cref{fig:diffuser_final_designs}, we present the final designs obtained by the classical method and the annealing-based optimization. Although the overall shape is very similar in both cases, the narrowing of the flow channels is slightly different. For the classical approach, the narrowing mainly happens immediately after the inlet and the channel is continued with an almost constant width. For the annealing-based approach, the narrowing stretches over the entire length of the channel. 
\Cref{fig:diffuserObjFrac} compares the optimization history for the objective function value $\objective$, i.e., the energy dissipation, and the volume fraction $\volumeFluid/|\domain|$ for both approaches.
First, we note the difference in the number of optimization steps, which is $19$ and $7$ for the classical and the annealing-based approach, respectively. This means a performance increase of about $63\%$ for the proposed approach. 
Next, we turn to the comparison of the final value of the objective function. Here, the classical approach performs a bit better than the annealing-based one. Specifically, the relative difference in $J$ is $6\%$.
Consider \Cref{tab:diffuserComparison} for more detailed numbers.
The comparison of the volume fraction shows no significant difference between the annealing solution and the unfiltered classical solution. The deviation of the volume fraction for the annealing-based approach from the target volume $\volumeLimit$ falls into the binary resolution range. The latter depicts the change in the volume fraction around $\volumeLimit$ resulting from flipping a single binary variable, i.e., adding or subtracting one element of solid material. For the classical solution, the volume fraction of the unfiltered solution satisfies the volume constraint, while filtering shifts it to a lower value, i.e., a smaller fluid region. This difference occurs because filtering to a strictly binary material distribution adds ``full'' material to elements that were previously in an intermediate state. 
\begin{figure}
    \centering
    % This file was created with tikzplotlib v0.10.1.
\begin{tikzpicture}

\definecolor{darkgray176}{RGB}{176,176,176}
\definecolor{gray}{RGB}{128,128,128}
\definecolor{lightgray204}{RGB}{204,204,204}

\begin{axis}[
legend cell align={left},
legend style={fill opacity=0.8, draw opacity=1, text opacity=1, draw=lightgray204},
tick align=outside,
tick pos=left,
title={Comparison Objective Function},
x grid style={darkgray176},
xlabel={Optimization Step},
xmin=0, xmax=19.9,
xtick style={color=black},
y grid style={darkgray176},
ylabel={Objective Function \(\displaystyle J\)},
ymin=0.00147010994227709, ymax=0.0315115099375456,
ytick style={color=black}
]
\addplot [semithick, gray, dashed, mark=x, mark size=3, mark options={solid}]
table {%
1 0.0028356281238802
2 0.0130741396783703
3 0.0288686511004581
4 0.0301459917559425
5 0.0198853466994459
6 0.0119825628596049
7 0.00808833710126787
8 0.00718151606397615
9 0.00693985586563039
10 0.00685910734995555
11 0.00683086316000876
12 0.0107308262120552
13 0.00924357490455159
14 0.00885689937841747
15 0.00877298804282824
16 0.00981034082417605
17 0.00947748297152956
18 0.00928752336606889
19 0.00919705379351396
};
\addlegendentry{Classical}
\addplot [semithick, gray, dash pattern=on 1pt off 3pt on 3pt off 3pt]
table {%
0 0.0095263488263556
19.9 0.0095263488263556
};
\addlegendentry{Classical (final, filtered)}
\addplot [semithick, black, mark=square*, mark size=3, mark options={solid}]
table {%
1 0.0253250454552656
2 0.0123584090452559
3 0.0113570345212802
4 0.0106745136808232
5 0.0102904540258453
6 0.0101424656070981
7 0.0100843715892529
};
\addlegendentry{Annealing}
\addplot [semithick, black]
table {%
0 0.0100843715892529
19.9 0.0100843715892529
};
\addlegendentry{Annealing (final)}
\end{axis}

\end{tikzpicture}
    % This file was created with tikzplotlib v0.10.1.
\begin{tikzpicture}

\definecolor{darkgray176}{RGB}{176,176,176}
\definecolor{gray}{RGB}{128,128,128}
\definecolor{lightgray}{RGB}{211,211,211}
\definecolor{lightgray204}{RGB}{204,204,204}

\begin{axis}[
legend cell align={left},
legend style={fill opacity=0.8, draw opacity=1, text opacity=1, draw=lightgray204},
tick align=outside,
tick pos=left,
title={Comparison Volume Fraction},
x grid style={darkgray176},
xlabel={Optimization Step},
xmin=0, xmax=19.9,
xtick style={color=black},
y grid style={darkgray176},
ylabel={Volume Fraction \(\displaystyle V_f/|\Omega|\)},
ymin=0.49, ymax=0.52,
ytick style={color=black}
]
\path [draw=lightgray, fill=lightgray, opacity=0.2]
(axis cs:0.1,0.5009765625)
--(axis cs:0.1,0.4990234375)
--(axis cs:0.3,0.4990234375)
--(axis cs:0.5,0.4990234375)
--(axis cs:0.7,0.4990234375)
--(axis cs:0.9,0.4990234375)
--(axis cs:1.1,0.4990234375)
--(axis cs:1.3,0.4990234375)
--(axis cs:1.5,0.4990234375)
--(axis cs:1.7,0.4990234375)
--(axis cs:1.9,0.4990234375)
--(axis cs:2.1,0.4990234375)
--(axis cs:2.3,0.4990234375)
--(axis cs:2.5,0.4990234375)
--(axis cs:2.7,0.4990234375)
--(axis cs:2.9,0.4990234375)
--(axis cs:3.1,0.4990234375)
--(axis cs:3.3,0.4990234375)
--(axis cs:3.5,0.4990234375)
--(axis cs:3.7,0.4990234375)
--(axis cs:3.9,0.4990234375)
--(axis cs:4.1,0.4990234375)
--(axis cs:4.3,0.4990234375)
--(axis cs:4.5,0.4990234375)
--(axis cs:4.7,0.4990234375)
--(axis cs:4.9,0.4990234375)
--(axis cs:5.1,0.4990234375)
--(axis cs:5.3,0.4990234375)
--(axis cs:5.5,0.4990234375)
--(axis cs:5.7,0.4990234375)
--(axis cs:5.9,0.4990234375)
--(axis cs:6.1,0.4990234375)
--(axis cs:6.3,0.4990234375)
--(axis cs:6.5,0.4990234375)
--(axis cs:6.7,0.4990234375)
--(axis cs:6.9,0.4990234375)
--(axis cs:7.1,0.4990234375)
--(axis cs:7.3,0.4990234375)
--(axis cs:7.5,0.4990234375)
--(axis cs:7.7,0.4990234375)
--(axis cs:7.9,0.4990234375)
--(axis cs:8.1,0.4990234375)
--(axis cs:8.3,0.4990234375)
--(axis cs:8.5,0.4990234375)
--(axis cs:8.7,0.4990234375)
--(axis cs:8.9,0.4990234375)
--(axis cs:9.1,0.4990234375)
--(axis cs:9.3,0.4990234375)
--(axis cs:9.5,0.4990234375)
--(axis cs:9.7,0.4990234375)
--(axis cs:9.9,0.4990234375)
--(axis cs:10.1,0.4990234375)
--(axis cs:10.3,0.4990234375)
--(axis cs:10.5,0.4990234375)
--(axis cs:10.7,0.4990234375)
--(axis cs:10.9,0.4990234375)
--(axis cs:11.1,0.4990234375)
--(axis cs:11.3,0.4990234375)
--(axis cs:11.5,0.4990234375)
--(axis cs:11.7,0.4990234375)
--(axis cs:11.9,0.4990234375)
--(axis cs:12.1,0.4990234375)
--(axis cs:12.3,0.4990234375)
--(axis cs:12.5,0.4990234375)
--(axis cs:12.7,0.4990234375)
--(axis cs:12.9,0.4990234375)
--(axis cs:13.1,0.4990234375)
--(axis cs:13.3,0.4990234375)
--(axis cs:13.5,0.4990234375)
--(axis cs:13.7,0.4990234375)
--(axis cs:13.9,0.4990234375)
--(axis cs:14.1,0.4990234375)
--(axis cs:14.3,0.4990234375)
--(axis cs:14.5,0.4990234375)
--(axis cs:14.7,0.4990234375)
--(axis cs:14.9,0.4990234375)
--(axis cs:15.1,0.4990234375)
--(axis cs:15.3,0.4990234375)
--(axis cs:15.5,0.4990234375)
--(axis cs:15.7,0.4990234375)
--(axis cs:15.9,0.4990234375)
--(axis cs:16.1,0.4990234375)
--(axis cs:16.3,0.4990234375)
--(axis cs:16.5,0.4990234375)
--(axis cs:16.7,0.4990234375)
--(axis cs:16.9,0.4990234375)
--(axis cs:17.1,0.4990234375)
--(axis cs:17.3,0.4990234375)
--(axis cs:17.5,0.4990234375)
--(axis cs:17.7,0.4990234375)
--(axis cs:17.9,0.4990234375)
--(axis cs:18.1,0.4990234375)
--(axis cs:18.3,0.4990234375)
--(axis cs:18.5,0.4990234375)
--(axis cs:18.7,0.4990234375)
--(axis cs:18.9,0.4990234375)
--(axis cs:19.1,0.4990234375)
--(axis cs:19.3,0.4990234375)
--(axis cs:19.5,0.4990234375)
--(axis cs:19.7,0.4990234375)
--(axis cs:19.9,0.4990234375)
--(axis cs:19.9,0.5009765625)
--(axis cs:19.9,0.5009765625)
--(axis cs:19.7,0.5009765625)
--(axis cs:19.5,0.5009765625)
--(axis cs:19.3,0.5009765625)
--(axis cs:19.1,0.5009765625)
--(axis cs:18.9,0.5009765625)
--(axis cs:18.7,0.5009765625)
--(axis cs:18.5,0.5009765625)
--(axis cs:18.3,0.5009765625)
--(axis cs:18.1,0.5009765625)
--(axis cs:17.9,0.5009765625)
--(axis cs:17.7,0.5009765625)
--(axis cs:17.5,0.5009765625)
--(axis cs:17.3,0.5009765625)
--(axis cs:17.1,0.5009765625)
--(axis cs:16.9,0.5009765625)
--(axis cs:16.7,0.5009765625)
--(axis cs:16.5,0.5009765625)
--(axis cs:16.3,0.5009765625)
--(axis cs:16.1,0.5009765625)
--(axis cs:15.9,0.5009765625)
--(axis cs:15.7,0.5009765625)
--(axis cs:15.5,0.5009765625)
--(axis cs:15.3,0.5009765625)
--(axis cs:15.1,0.5009765625)
--(axis cs:14.9,0.5009765625)
--(axis cs:14.7,0.5009765625)
--(axis cs:14.5,0.5009765625)
--(axis cs:14.3,0.5009765625)
--(axis cs:14.1,0.5009765625)
--(axis cs:13.9,0.5009765625)
--(axis cs:13.7,0.5009765625)
--(axis cs:13.5,0.5009765625)
--(axis cs:13.3,0.5009765625)
--(axis cs:13.1,0.5009765625)
--(axis cs:12.9,0.5009765625)
--(axis cs:12.7,0.5009765625)
--(axis cs:12.5,0.5009765625)
--(axis cs:12.3,0.5009765625)
--(axis cs:12.1,0.5009765625)
--(axis cs:11.9,0.5009765625)
--(axis cs:11.7,0.5009765625)
--(axis cs:11.5,0.5009765625)
--(axis cs:11.3,0.5009765625)
--(axis cs:11.1,0.5009765625)
--(axis cs:10.9,0.5009765625)
--(axis cs:10.7,0.5009765625)
--(axis cs:10.5,0.5009765625)
--(axis cs:10.3,0.5009765625)
--(axis cs:10.1,0.5009765625)
--(axis cs:9.9,0.5009765625)
--(axis cs:9.7,0.5009765625)
--(axis cs:9.5,0.5009765625)
--(axis cs:9.3,0.5009765625)
--(axis cs:9.1,0.5009765625)
--(axis cs:8.9,0.5009765625)
--(axis cs:8.7,0.5009765625)
--(axis cs:8.5,0.5009765625)
--(axis cs:8.3,0.5009765625)
--(axis cs:8.1,0.5009765625)
--(axis cs:7.9,0.5009765625)
--(axis cs:7.7,0.5009765625)
--(axis cs:7.5,0.5009765625)
--(axis cs:7.3,0.5009765625)
--(axis cs:7.1,0.5009765625)
--(axis cs:6.9,0.5009765625)
--(axis cs:6.7,0.5009765625)
--(axis cs:6.5,0.5009765625)
--(axis cs:6.3,0.5009765625)
--(axis cs:6.1,0.5009765625)
--(axis cs:5.9,0.5009765625)
--(axis cs:5.7,0.5009765625)
--(axis cs:5.5,0.5009765625)
--(axis cs:5.3,0.5009765625)
--(axis cs:5.1,0.5009765625)
--(axis cs:4.9,0.5009765625)
--(axis cs:4.7,0.5009765625)
--(axis cs:4.5,0.5009765625)
--(axis cs:4.3,0.5009765625)
--(axis cs:4.1,0.5009765625)
--(axis cs:3.9,0.5009765625)
--(axis cs:3.7,0.5009765625)
--(axis cs:3.5,0.5009765625)
--(axis cs:3.3,0.5009765625)
--(axis cs:3.1,0.5009765625)
--(axis cs:2.9,0.5009765625)
--(axis cs:2.7,0.5009765625)
--(axis cs:2.5,0.5009765625)
--(axis cs:2.3,0.5009765625)
--(axis cs:2.1,0.5009765625)
--(axis cs:1.9,0.5009765625)
--(axis cs:1.7,0.5009765625)
--(axis cs:1.5,0.5009765625)
--(axis cs:1.3,0.5009765625)
--(axis cs:1.1,0.5009765625)
--(axis cs:0.9,0.5009765625)
--(axis cs:0.7,0.5009765625)
--(axis cs:0.5,0.5009765625)
--(axis cs:0.3,0.5009765625)
--(axis cs:0.1,0.5009765625)
--cycle;
\addlegendimage{area legend, draw=lightgray, fill=lightgray, opacity=0.2}
\addlegendentry{Binary Resolution}

\addplot [semithick, black, dotted]
table {%
0 0.5
19.9 0.5
};
\addlegendentry{$V_{\mathrm{max}}$}
\addplot [semithick, lightgray, forget plot]
table {%
0 0.5009765625
19.9 0.5009765625
};
\addplot [semithick, lightgray, forget plot]
table {%
0 0.4990234375
19.9 0.4990234375
};
\addplot [semithick, gray, dashed, mark=x, mark size=3, mark options={solid}]
table {%
1 0.8
2 0.6
3 0.499989219033465
4 0.500060363597868
5 0.500011010448369
6 0.499917531192211
7 0.500077014869896
8 0.499945280509116
9 0.49999481202698
10 0.499994950315845
11 0.499988577673993
12 0.499996091633121
13 0.500002956909401
14 0.500000596249279
15 0.499990044013989
16 0.499998153540262
17 0.499997710184589
18 0.499999898925017
19 0.499997821479091
};
\addlegendentry{Classical}
\addplot [semithick, gray, dash pattern=on 1pt off 3pt on 3pt off 3pt]
table {%
0 0.4931640625
19.9 0.4931640625
};
\addlegendentry{Classical (filtered)}
\addplot [semithick, black, mark=square*, mark size=3, mark options={solid}]
table {%
1 0.513671875
2 0.501953125
3 0.5009765625
4 0.5009765625
5 0.5009765625
6 0.5009765625
7 0.5009765625
};
\addlegendentry{Annealing}
\end{axis}

\end{tikzpicture}
    \caption{Diffuser problem: optimization history for the objective function $\objective$ (left) and the volume fraction $\volumeFluid/|\domain|$ (right).
    For the volume fraction, \textit{Binary Resolution} depicts the change in the volume fraction around $\volumeLimit$ by adding or subtracting one element of solid material.
    }
    \label{fig:diffuserObjFrac}
\end{figure}
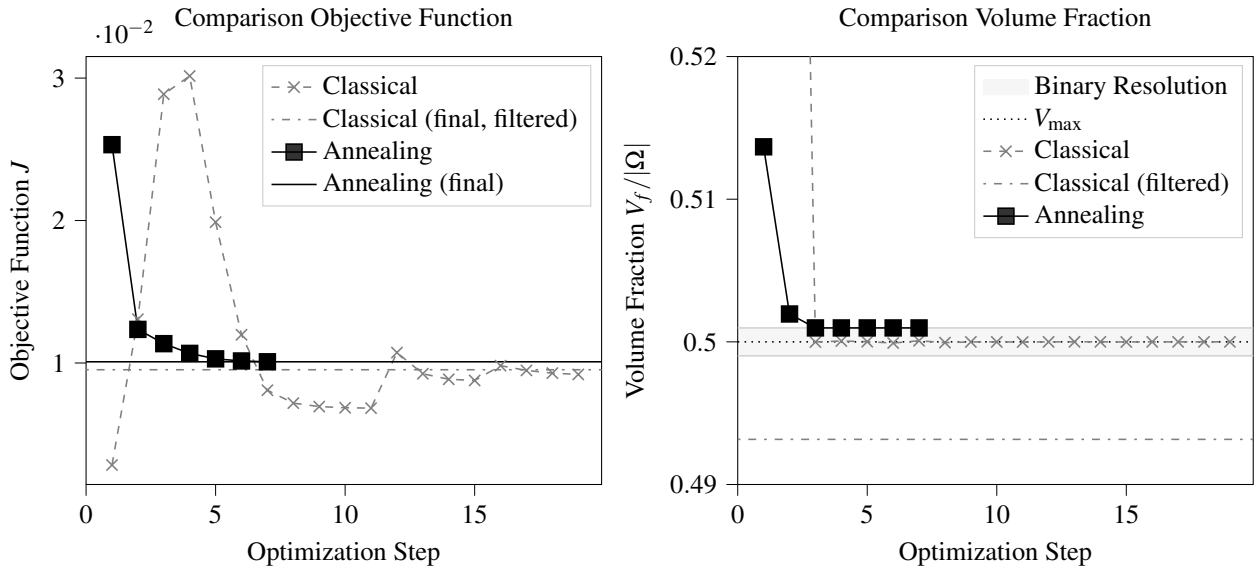
\begin{table}
    \centering
    \begin{tabular}{l|ccc}
        \textbf{Quantity} & \textbf{Classical} & \textbf{Annealing} & \textbf{Relative Difference} \\
        \hline
        Number of Opt. Steps $\nopt$& $\nopt_C = 19$ & $\nopt_A = 7$ & $\frac{\nopt_A - \nopt_C}{\nopt_C} = -63.2\%$ \\ 
        Objective Function $\objective$ & $\objective_C = 0.953\times10^{-2}$ & $\objective_A = 1.01\times10^{-2}$ & $\frac{\objective_A - \objective_C}{\objective_{C}}=+5.86\%$         
    \end{tabular}
    \caption{Diffuser problem: comparison of the number of optimization steps $\nopt$ and the final objective function value $\objective$.}
    \label{tab:diffuserComparison}
\end{table}
\bigskip\par
Next, we will report the results of the double pipe problem. Here, we set $\volumeLimit=\frac{1}{3}|\domain|$ and choose material properties and \gls{qubo} hyperparameters as $\viscosity=1.0$, $\resistanceCoeffSolid=12.5$, and $\paramVolume=0.05$, respectively. For the time-out time of the \gls{ae}, we use $\timeOut = 10,000\,ms$. 
\par
\begin{figure}
    \centering
    \resizebox{0.45\textwidth}{!}{
        % This file was created with tikzplotlib v0.10.1.
\begin{tikzpicture}

\definecolor{darkgray176}{RGB}{176,176,176}

\begin{axis}[
tick align=outside,
tick pos=left,
title={Classical},
x grid style={darkgray176},
xlabel={$x$},
xmin=0, xmax=1.5,
xtick style={color=black},
y grid style={darkgray176},
ylabel={$y$},
ymin=0, ymax=1,
ytick style={color=black}
]
\addplot graphics [includegraphics cmd=\pgfimage,xmin=0, xmax=1.5, ymin=0, ymax=1] {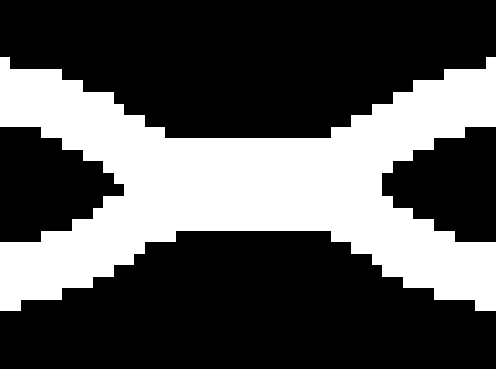};
\end{axis}

\end{tikzpicture}
    }
    \hfill
    \resizebox{0.45\textwidth}{!}{
        % This file was created with tikzplotlib v0.10.1.
\begin{tikzpicture}

\definecolor{darkgray176}{RGB}{176,176,176}

\begin{axis}[
tick align=outside,
tick pos=left,
title={Annealing},
x grid style={darkgray176},
xlabel={$x$},
xmin=0, xmax=1.5,
xtick style={color=black},
y grid style={darkgray176},
ylabel={$y$},
ymin=0, ymax=1,
ytick style={color=black}
]
\addplot graphics [includegraphics cmd=\pgfimage,xmin=0, xmax=1.5, ymin=0, ymax=1] {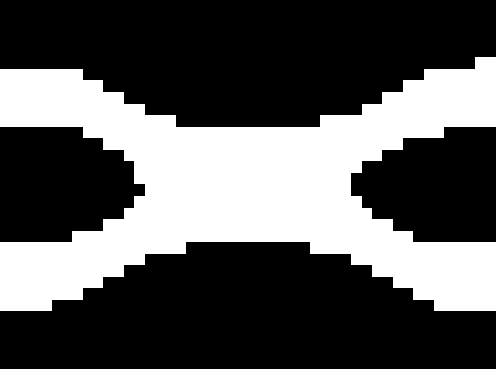};
\end{axis}

\end{tikzpicture}
    }

    \caption{Double pipe problem: final designs for the classical approach (left) and the annealing approach (right).
    }
    \label{fig:double_pipe_final_designs}
\end{figure}

\begin{figure}
    \centering
    % This file was created with tikzplotlib v0.10.1.
\begin{tikzpicture}

\definecolor{darkgray176}{RGB}{176,176,176}
\definecolor{gray}{RGB}{128,128,128}
\definecolor{lightgray204}{RGB}{204,204,204}

\begin{axis}[
legend cell align={left},
legend style={fill opacity=0.8, draw opacity=1, text opacity=1, draw=lightgray204},
tick align=outside,
tick pos=left,
title={Comparison Objective Function},
x grid style={darkgray176},
xlabel={Optimization Step},
xmin=-0.45, xmax=31.45,
xtick style={color=black},
y grid style={darkgray176},
ylabel={Objective Function \(\displaystyle J\)},
ymin=0.000338399104452715, ymax=0.0212995155447755,
ytick style={color=black}
]
\addplot [semithick, gray, dashed, mark=x, mark size=3, mark options={solid}]
table {%
1 0.00129117712446739
2 0.00130800413403712
3 0.00133599023701187
4 0.00139175383369178
5 0.00138975378070405
6 0.00206738109245555
7 0.00206474459267706
8 0.00826581351103336
9 0.00792100023818086
10 0.00754235862135257
11 0.00718997209300769
12 0.00680110417355453
13 0.00640917742275954
14 0.00605243283877356
15 0.00580164145607527
16 0.00562977265371694
17 0.00552629965012149
18 0.00543773258163264
19 0.00537862976464335
20 0.00533253542341234
21 0.0125935519959001
22 0.00951373746604438
23 0.00887727905062749
24 0.00872037446219812
25 0.00864750721520196
26 0.0108686285711545
27 0.0101095329097123
28 0.00976272946033845
29 0.00961873874934866
30 0.00955430827338161
};
\addlegendentry{Classical}
\addplot [semithick, gray, dash pattern=on 1pt off 3pt on 3pt off 3pt]
table {%
-0.45 0.00998718787665346
31.45 0.00998718787665346
};
\addlegendentry{Classical (final, filtered)}
\addplot [semithick, black, mark=square*, mark size=3, mark options={solid}]
table {%
1 0.0203467375247609
2 0.014870822288675
3 0.0134055065240296
4 0.0126872554151058
5 0.0122290171790956
6 0.0120896431674815
7 0.011983384937855
};
\addlegendentry{Annealing}
\addplot [semithick, black]
table {%
-0.45 0.011983384937855
31.45 0.011983384937855
};
\addlegendentry{Annealing (final)}
\end{axis}

\end{tikzpicture}
    \input{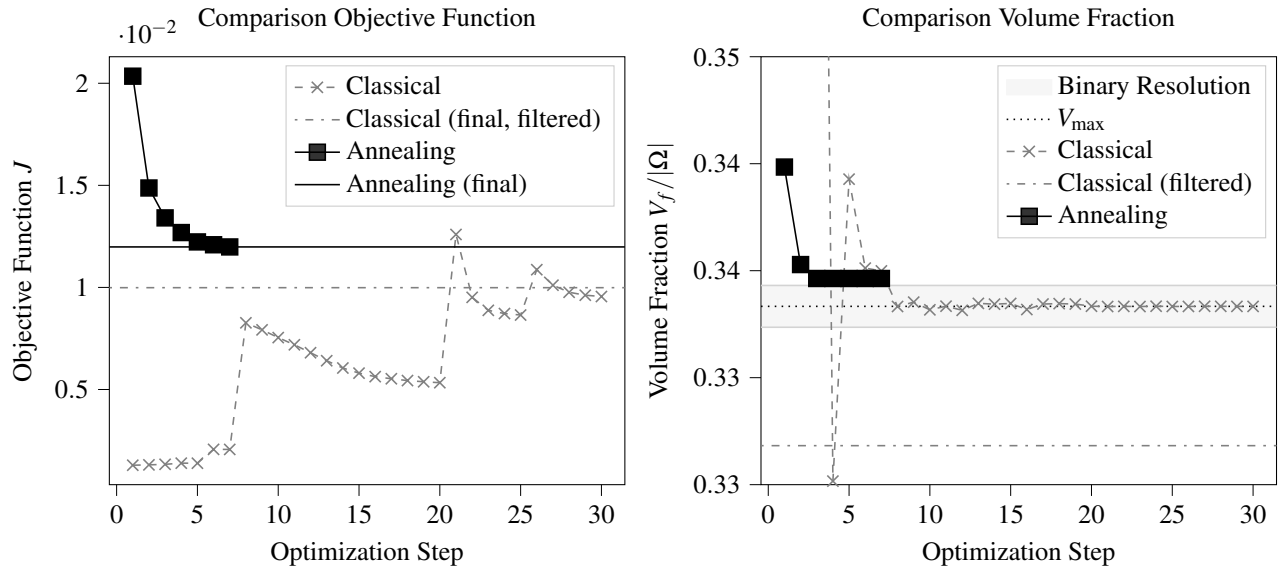}
    \caption{Double pipe problem: optimization history for the objective function $\objective$ (left) and the volume fraction $\volumeFluid/|\domain|$ (right).
    For the volume fraction, \textit{Binary Resolution} depicts the change in the volume fraction around $\volumeLimit$ by adding or subtracting one element of solid material.
    }
    \label{fig:doublePipeObjFrac}
\end{figure}

 
The final designs are shown in \Cref{fig:double_pipe_final_designs}. Again, the overall shape is very similar. In both cases, two flow channels from the inlet merge in the center of the domain before splitting again toward the outlets. There are slight differences in the curvature near the inlet and outlet. Most noticeable is the difference in the width of the merged channel in the center, which is greater for the annealing-based result.
The comparison of the optimization history is provided in \Cref{fig:doublePipeObjFrac} and \Cref{tab:doublePipeComparison}. Similar to the diffuser test case, the annealing-based approach needs fewer ($7$) optimization steps than the classical approach ($30$). The time intensity is thus decreased by around $77\%$. For the final values of the objective function, a greater gap than for the diffuser case exists. The relative difference is $20\%$, representing a substantial deviation with respect to the energy dissipation of the designs.
Nevertheless, the comparison of the volume fraction does not show relevant differences between the two approaches.
\begin{table}
    \centering
    \begin{tabular}{l|ccc}
        \textbf{Quantity} & \textbf{Classical} & \textbf{Annealing} & \textbf{Relative Difference} \\
        \hline
        Number of Opt. Steps $\nopt$& $\nopt_C = 30$ & $\nopt_A = 7$ & $\frac{\nopt_A - \nopt_C}{\nopt_C} = -76.7\%$ \\ 
        Objective Function $\objective$ & $\objective_C = 0.999\times10^{-2}$ & $\objective_A = 1.20\times10^{-2}$ & $\frac{\objective_A - \objective_C}{\objective_{C}}=+20.0\%$         
    \end{tabular}
    \caption{Double pipe problem: comparison of the number of optimization steps $\nopt$ and the final objective function value $\objective$.}
    \label{tab:doublePipeComparison}
\end{table}
\bigskip\par
Summarizing the results of the comparison performed in the above test cases, we can state that the proposed approach requires far fewer optimization steps to converge and deliver qualitatively comparable designs. However, the performance of the design found with the annealing-based update strategy is around 6\% and 20\% worse for the diffuser and double pipe problem, respectively. The optimization history suggests that the annealing-based approach settles in one of the many local optima faster than the classical approach. In this sense, the investigated approach is less exploratory, and consequently, the final design depends noticeably on the initial design and the corresponding flow field. An explanation can be found in the type of design update strategy we follow in this work. In contrast to gradient-based approaches, our strategy ignores any feedback mechanism, i.e., the effect of the design change on the flow field. Instead, it only considers what is optimal with respect to the current flow field. This can be described as \textit{intermediate optimality} per optimization step. Furthermore, the pure binary character of our design description differs from the continuous description in classical approaches and shrinks the space of possibilities of the design. Thus, it can be expected that the resulting designs show differences in their performance. In conclusion, the presented way of integrating an Ising machine formulation to compute design updates based on an intermediate optimality criterion in an iterative scheme shows the potential of accelerating the optimization process but is limited in its ability to deliver high-performance designs.

%%%%%%%%%%%%%%%%
%% CONCLUSION %%
%%%%%%%%%%%%%%%%
\section*{Conclusion}
\label{sec:conclusion}
% Approach.
In this work, we developed a novel Ising machine formulation, specifically a \gls{qubo} problem, for computing design updates in the topology optimization of flow channels with the goal of minimizing energy dissipation under a volume constraint. We integrated this formulation into a classical two-step optimization approach, where the flow field and the design update are computed sequentially. Specifically for our approach, the design update was obtained by solving the \gls{qubo} problem. Our objective was then to analyze the proposed formulation and investigate whether the proposed formulation can enhance the performance of the optimization approach.
\par
% Results
To address these questions, we performed numerical experiments for two common benchmark test cases.
After studying the influence of different contributions to the objective function of the \gls{qubo} problem, we found that the regularization term typically used in level-set-based topology optimization for structural mechanics is not essential when applied to the design optimization of flow channels, which is in agreement with the existing literature.
During the comparison with a classical method, we observed that our scheme can significantly reduce the number of optimization steps required to achieve a design quality comparable to that of the reference method, thereby accelerating the overall optimization procedure.
However, the resulting performance of the designs, as measured by the value of the objective function, was inferior to that of the traditional approach. This was attributed to the limited exploratory capability of the proposed approach, which determines design updates solely based on intermediate optimality without incorporating sensitivity information.
\par
% Limitations
There are several limitations to our work that must be acknowledged. For example, the findings are based solely on one type of Ising machine, namely the GPU-based \gls{ae}, which may not fully capture the performance characteristics of other Ising machine types such as \gls{qa}.
Additionally, the study is limited by its use of only two relatively simple 2D test cases, which may not fully represent the complexity and diversity of real-world flow optimization problems.
\par
% Outlook
Nevertheless, the results encourage further exploration of using Ising machines for topology optimization in flow problems and this work provides valuable insights into designing and integrating \gls{qubo} problems for different engineering tasks.
In addition, it may be worthwhile to extend the Ising machine formulation for the design update to the entire optimization problem, including the governing equations. This would eliminate the need for an iterative two-step optimization approach and, thereby, resolve the issue of intermediate optimality. As a result, such an approach is expected to be significantly more efficient and to yield designs with superior performance compared to classical methods. Regarding the integration of the governing equations, we note that \gls{qa} has already been used for solving a reduced version of the Navier-Stokes equations~\cite{Ray2022}. Thus, this presents an intriguing avenue for future research.

%%%%%%%%%%%%%%%%%%%%%%%
%% DATA AVAILABILITY %%
%%%%%%%%%%%%%%%%%%%%%%%
\section*{Data Availability}
The datasets generated and analyzed during the current study are available in the \texttt{TopoFlow} repository on \textit{GitHub}~\cite{Key2024TopoFlow}.
% The datasets generated and analyzed during the current study are available from the corresponding author upon reasonable request.

%%%%%%%%%%%%%%%%%%%%%%
%% ACKNOWLEDGEMENTS %%
%%%%%%%%%%%%%%%%%%%%%%
\section*{Acknowledgements}
This work was partially supported by JSPS KAKENHI (Grant Number JP23H05447), the Council for Science, Technology, and Innovation (CSTI) through the Cross-ministerial Strategic Innovation Promotion Program (SIP), ``Promoting the application of advanced quantum technology platforms to social issues'' (Funding agency: QST), JST (Grant Number JPMJPF2221). One of the authors S. T. wishes to express their gratitude to the World Premier International Research Center Initiative (WPI), MEXT, Japan, for their support of the Human Biology-Microbiome-Quantum Research Center (Bio2Q).
Finally, we would like to express our gratitude to Norbert Hosters for supporting Shiori Aoki during her stay at the Chair for Computational Analysis of Technical Systems, RWTH Aachen University, which greatly contributed to the success of this research.

%%%%%%%%%%%%%%%%%%%%%%%%%%
%% AUTHOR CONTRIBUTIONS %%
%%%%%%%%%%%%%%%%%%%%%%%%%%
\section*{Author contributions}
Y.S., F.K., and M.M. designed the research.
S.A. mainly performed numerical simulations with the help of F.K. and Y.S..
Y.S. and F.K. wrote the first manuscript.
F.K., K.E. Y.M., S.T., M.B., and M.M. supervised the project.
All authors analyzed the obtained results and contributed to the improvement of the manuscript.

% \section*{Additional information}

%%%%%%%%%%%%%%%%%%%%%%%%%
%% COMPETING INTERESTS %%
%%%%%%%%%%%%%%%%%%%%%%%%%
\section*{Competing interests} 
The authors declare the following potential conflict of interest: Y.M. is affiliated with Fixstars Corporation, which provided access to the Fixstars Amplify AE used for calculations in this study. However, this affiliation did not influence the interpretation of the results or the conclusions drawn from this research. All other authors declare no conflict of interest.

%%%%%%%%%%%%%%%%
%% REFERENCES %%
%%%%%%%%%%%%%%%%
\bibliography{references}

\begin{thebibliography}{10}
\urlstyle{rm}
\expandafter\ifx\csname url\endcsname\relax
  \def\url#1{\texttt{#1}}\fi
\expandafter\ifx\csname urlprefix\endcsname\relax\def\urlprefix{URL }\fi
\expandafter\ifx\csname doiprefix\endcsname\relax\def\doiprefix{DOI: }\fi
\providecommand{\bibinfo}[2]{#2}
\providecommand{\eprint}[2][]{\url{#2}}

\bibitem{Mohseni2022}
\bibinfo{author}{Mohseni, N.}, \bibinfo{author}{McMahon, P.~L.} \& \bibinfo{author}{Byrnes, T.}
\newblock \bibinfo{journal}{\bibinfo{title}{Ising machines as hardware solvers of combinatorial optimization problems}}.
\newblock {\emph{\JournalTitle{Nature Reviews Physics}}} \textbf{\bibinfo{volume}{4}}, \bibinfo{pages}{363--379} (\bibinfo{year}{2022}).

\bibitem{Apolloni1989}
\bibinfo{author}{Apolloni, B.}, \bibinfo{author}{Carvalho, C.} \& \bibinfo{author}{De~Falco, D.}
\newblock \bibinfo{journal}{\bibinfo{title}{Quantum stochastic optimization}}.
\newblock {\emph{\JournalTitle{Stochastic Processes and their Applications}}} \textbf{\bibinfo{volume}{33}}, \bibinfo{pages}{233--244} (\bibinfo{year}{1989}).

\bibitem{Finnila1994}
\bibinfo{author}{Finnila, A.~B.}, \bibinfo{author}{Gomez, M.~A.}, \bibinfo{author}{Sebenik, C.}, \bibinfo{author}{Stenson, C.} \& \bibinfo{author}{Doll, J.~D.}
\newblock \bibinfo{journal}{\bibinfo{title}{{Quantum annealing: A new method for minimizing multidimensional functions}}}.
\newblock {\emph{\JournalTitle{Chemical Physics Letters}}} \textbf{\bibinfo{volume}{219}}, \bibinfo{pages}{343--348} (\bibinfo{year}{1994}).

\bibitem{Kadowaki1998}
\bibinfo{author}{Kadowaki, T.} \& \bibinfo{author}{Nishimori, H.}
\newblock \bibinfo{journal}{\bibinfo{title}{Quantum annealing in the transverse ising model}}.
\newblock {\emph{\JournalTitle{Physical Review E}}} \textbf{\bibinfo{volume}{58}}, \bibinfo{pages}{5355} (\bibinfo{year}{1998}).

\bibitem{Tanaka2017}
\bibinfo{author}{Tanaka, S.}, \bibinfo{author}{Tamura, R.} \& \bibinfo{author}{Chakrabarti, B.~K.}
\newblock \emph{\bibinfo{title}{Quantum spin glasses, annealing and computation}} (\bibinfo{publisher}{Cambridge University Press}, \bibinfo{year}{2017}).

\bibitem{Albash2018}
\bibinfo{author}{Albash, T.} \& \bibinfo{author}{Lidar, D.~A.}
\newblock \bibinfo{journal}{\bibinfo{title}{Adiabatic quantum computation}}.
\newblock {\emph{\JournalTitle{Reviews of Modern Physics}}} \textbf{\bibinfo{volume}{90}}, \bibinfo{pages}{015002} (\bibinfo{year}{2018}).

\bibitem{Hauke2020}
\bibinfo{author}{Hauke, P.}, \bibinfo{author}{Katzgraber, H.~G.}, \bibinfo{author}{Lechner, W.}, \bibinfo{author}{Nishimori, H.} \& \bibinfo{author}{Oliver, W.~D.}
\newblock \bibinfo{journal}{\bibinfo{title}{Perspectives of quantum annealing: Methods and implementations}}.
\newblock {\emph{\JournalTitle{Reports on Progress in Physics}}} \textbf{\bibinfo{volume}{83}}, \bibinfo{pages}{054401} (\bibinfo{year}{2020}).

\bibitem{Yarkoni2022}
\bibinfo{author}{Yarkoni, S.}, \bibinfo{author}{Raponi, E.}, \bibinfo{author}{B{\"a}ck, T.} \& \bibinfo{author}{Schmitt, S.}
\newblock \bibinfo{journal}{\bibinfo{title}{Quantum annealing for industry applications: Introduction and review}}.
\newblock {\emph{\JournalTitle{Reports on Progress in Physics}}} \textbf{\bibinfo{volume}{85}}, \bibinfo{pages}{104001} (\bibinfo{year}{2022}).

\bibitem{Inagaki2016}
\bibinfo{author}{Inagaki, T.} \emph{et~al.}
\newblock \bibinfo{journal}{\bibinfo{title}{{A coherent Ising machine for 2000-node optimization problems}}}.
\newblock {\emph{\JournalTitle{Science}}} \textbf{\bibinfo{volume}{354}}, \bibinfo{pages}{603--606} (\bibinfo{year}{2016}).

\bibitem{Goto2019}
\bibinfo{author}{Goto, H.}, \bibinfo{author}{Tatsumura, K.} \& \bibinfo{author}{Dixon, A.~R.}
\newblock \bibinfo{journal}{\bibinfo{title}{{Combinatorial optimization by simulating adiabatic bifurcations in nonlinear Hamiltonian systems}}}.
\newblock {\emph{\JournalTitle{Science Advances}}} \textbf{\bibinfo{volume}{5}}, \bibinfo{pages}{eaav2372} (\bibinfo{year}{2019}).

\bibitem{Matsubara2020}
\bibinfo{author}{Matsubara, S.} \emph{et~al.}
\newblock \bibinfo{title}{{Digital Annealer for High-Speed Solving of Combinatorial optimization Problems and Its Applications}}.
\newblock In \emph{\bibinfo{booktitle}{2020 25th Asia and South Pacific Design Automation Conference (ASP-DAC)}}, \bibinfo{pages}{667--672} (\bibinfo{year}{2020}).

\bibitem{Goto2021}
\bibinfo{author}{Goto, H.} \emph{et~al.}
\newblock \bibinfo{journal}{\bibinfo{title}{{High-performance combinatorial optimization based on classical mechanics}}}.
\newblock {\emph{\JournalTitle{Science Advances}}} \textbf{\bibinfo{volume}{7}}, \bibinfo{pages}{eabe7953} (\bibinfo{year}{2021}).

\bibitem{Tatsumura2021}
\bibinfo{author}{Tatsumura, K.}, \bibinfo{author}{Yamasaki, M.} \& \bibinfo{author}{Goto, H.}
\newblock \bibinfo{journal}{\bibinfo{title}{{Scaling out Ising machines using a multi-chip architecture for simulated bifurcation}}}.
\newblock {\emph{\JournalTitle{Nature Electronics}}} \textbf{\bibinfo{volume}{4}}, \bibinfo{pages}{208--217} (\bibinfo{year}{2021}).

\bibitem{Neukart2017}
\bibinfo{author}{Neukart, F.} \emph{et~al.}
\newblock \bibinfo{journal}{\bibinfo{title}{Traffic flow optimization using a quantum annealer}}.
\newblock {\emph{\JournalTitle{Frontiers in ICT}}} \textbf{\bibinfo{volume}{4}}, \bibinfo{pages}{29} (\bibinfo{year}{2017}).

\bibitem{Ohzeki2019}
\bibinfo{author}{Ohzeki, M.}, \bibinfo{author}{Miki, A.}, \bibinfo{author}{Miyama, M.~J.} \& \bibinfo{author}{Terabe, M.}
\newblock \bibinfo{journal}{\bibinfo{title}{Control of automated guided vehicles without collision by quantum annealer and digital devices}}.
\newblock {\emph{\JournalTitle{Frontiers in Computer Science}}} \textbf{\bibinfo{volume}{1}}, \bibinfo{pages}{9} (\bibinfo{year}{2019}).

\bibitem{Stollenwerk2019}
\bibinfo{author}{Stollenwerk, T.} \emph{et~al.}
\newblock \bibinfo{journal}{\bibinfo{title}{Quantum annealing applied to de-conflicting optimal trajectories for air traffic management}}.
\newblock {\emph{\JournalTitle{IEEE Transactions on Intelligent Transportation Systems}}} \textbf{\bibinfo{volume}{21}}, \bibinfo{pages}{285--297} (\bibinfo{year}{2019}).

\bibitem{Kanai2024}
\bibinfo{author}{Kanai, H.}, \bibinfo{author}{Yamashita, M.}, \bibinfo{author}{Tanahashi, K.} \& \bibinfo{author}{Tanaka, S.}
\newblock \bibinfo{journal}{\bibinfo{title}{{Annealing-Assisted Column Generation for Inequality-Constrained Combinatorial optimization Problems}}}.
\newblock {\emph{\JournalTitle{IEEE Access}}} \textbf{\bibinfo{volume}{12}}, \bibinfo{pages}{157669--157685} (\bibinfo{year}{2024}).

\bibitem{Rieffel2015}
\bibinfo{author}{Rieffel, E.~G.} \emph{et~al.}
\newblock \bibinfo{journal}{\bibinfo{title}{A case study in programming a quantum annealer for hard operational planning problems}}.
\newblock {\emph{\JournalTitle{Quantum Information Processing}}} \textbf{\bibinfo{volume}{14}}, \bibinfo{pages}{1--36} (\bibinfo{year}{2015}).

\bibitem{Rosenberg2016}
\bibinfo{author}{Rosenberg, G.} \emph{et~al.}
\newblock \bibinfo{journal}{\bibinfo{title}{Solving the optimal trading trajectory problem using a quantum annealer}}.
\newblock {\emph{\JournalTitle{IEEE Journal of Selected Topics in Signal Processing}}} \textbf{\bibinfo{volume}{10}}, \bibinfo{pages}{1053--1060} (\bibinfo{year}{2016}).

\bibitem{Venturelli2019}
\bibinfo{author}{Venturelli, D.} \& \bibinfo{author}{Kondratyev, A.}
\newblock \bibinfo{journal}{\bibinfo{title}{Reverse quantum annealing approach to portfolio optimization problems}}.
\newblock {\emph{\JournalTitle{Quantum Machine Intelligence}}} \textbf{\bibinfo{volume}{1}}, \bibinfo{pages}{17--30} (\bibinfo{year}{2019}).

\bibitem{Vreumingen2019}
\bibinfo{author}{van Vreumingen, D.} \emph{et~al.}
\newblock \bibinfo{title}{Quantum-assisted finite-element design optimization.}, \doiprefix\url{10.48550/arXiv.1908.03947} (\bibinfo{year}{2019}).

\bibitem{Raisuddin2022}
\bibinfo{author}{Raisuddin, O.~M.} \& \bibinfo{author}{De, S.}
\newblock \bibinfo{journal}{\bibinfo{title}{{FEqa: Finite element computations on quantum annealers}}}.
\newblock {\emph{\JournalTitle{Computer Methods in Applied Mechanics and Engineering}}} \textbf{\bibinfo{volume}{395}}, \bibinfo{pages}{115014} (\bibinfo{year}{2022}).

\bibitem{Endo2022}
\bibinfo{author}{Endo, K.}, \bibinfo{author}{Matsuda, Y.}, \bibinfo{author}{Tanaka, S.} \& \bibinfo{author}{Muramatsu, M.}
\newblock \bibinfo{journal}{\bibinfo{title}{A phase-field model by an ising machine and its application to the phase-separation structure of a diblock polymer}}.
\newblock {\emph{\JournalTitle{Scientific Reports}}} \textbf{\bibinfo{volume}{12}}, \bibinfo{pages}{10794} (\bibinfo{year}{2022}).

\bibitem{Ray2022}
\bibinfo{author}{Ray, N.}, \bibinfo{author}{Banerjee, T.}, \bibinfo{author}{Nadiga, B.} \& \bibinfo{author}{Karra, S.}
\newblock \bibinfo{journal}{\bibinfo{title}{On the viability of quantum annealers to solve fluid flows}}.
\newblock {\emph{\JournalTitle{Frontiers in Mechanical Engineering}}} \textbf{\bibinfo{volume}{8}}, \bibinfo{pages}{906696} (\bibinfo{year}{2022}).

\bibitem{Ye2023}
\bibinfo{author}{Ye, Z.}, \bibinfo{author}{Qian, X.} \& \bibinfo{author}{Pan, W.}
\newblock \bibinfo{journal}{\bibinfo{title}{Quantum topology optimization via quantum annealing}}.
\newblock {\emph{\JournalTitle{IEEE Transactions on Quantum Engineering}}} \textbf{\bibinfo{volume}{4}}, \bibinfo{pages}{1--15} (\bibinfo{year}{2023}).

\bibitem{Honda2024}
\bibinfo{author}{Honda, R.} \emph{et~al.}
\newblock \bibinfo{journal}{\bibinfo{title}{Development of optimization method for truss structure by quantum annealing}}.
\newblock {\emph{\JournalTitle{Scientific Reports}}} \textbf{\bibinfo{volume}{14}}, \bibinfo{pages}{13872} (\bibinfo{year}{2024}).

\bibitem{Wang2024}
\bibinfo{author}{Wang, X.}, \bibinfo{author}{Wang, Z.} \& \bibinfo{author}{Ni, B.}
\newblock \bibinfo{journal}{\bibinfo{title}{Mapping structural topology optimization problems to quantum annealing}}.
\newblock {\emph{\JournalTitle{Structural and Multidisciplinary Optimization}}} \textbf{\bibinfo{volume}{67}}, \bibinfo{pages}{74} (\bibinfo{year}{2024}).

\bibitem{Maruo2022}
\bibinfo{author}{Maruo, A.}, \bibinfo{author}{Soeda, T.} \& \bibinfo{author}{Igarashi, H.}
\newblock \bibinfo{journal}{\bibinfo{title}{Topology optimization of electromagnetic devices using digital annealer}}.
\newblock {\emph{\JournalTitle{IEEE Transactions on Magnetics}}} \textbf{\bibinfo{volume}{58}}, \bibinfo{pages}{1--4} (\bibinfo{year}{2022}).

\bibitem{Key2024}
\bibinfo{author}{Key, F.} \& \bibinfo{author}{Freinberger, L.}
\newblock \bibinfo{journal}{\bibinfo{title}{A formulation of structural design optimization problems for quantum annealing}}.
\newblock {\emph{\JournalTitle{Mathematics}}} \textbf{\bibinfo{volume}{12}}, \bibinfo{pages}{482} (\bibinfo{year}{2024}).

\bibitem{Okada2023}
\bibinfo{author}{Okada, A.} \emph{et~al.}
\newblock \bibinfo{journal}{\bibinfo{title}{Design optimization of noise filter using quantum annealer}}.
\newblock {\emph{\JournalTitle{IEEE Access}}} \textbf{\bibinfo{volume}{11}}, \bibinfo{pages}{44343--44349} (\bibinfo{year}{2023}).

\bibitem{Inoue2022}
\bibinfo{author}{Inoue, T.} \emph{et~al.}
\newblock \bibinfo{journal}{\bibinfo{title}{Towards optimization of photonic-crystal surface-emitting lasers via quantum annealing}}.
\newblock {\emph{\JournalTitle{Optics Express}}} \textbf{\bibinfo{volume}{30}}, \bibinfo{pages}{43503--43512} (\bibinfo{year}{2022}).

\bibitem{Matsumori2022}
\bibinfo{author}{Matsumori, T.}, \bibinfo{author}{Taki, M.} \& \bibinfo{author}{Kadowaki, T.}
\newblock \bibinfo{journal}{\bibinfo{title}{Application of qubo solver using black-box optimization to structural design for resonance avoidance}}.
\newblock {\emph{\JournalTitle{Scientific Reports}}} \textbf{\bibinfo{volume}{12}}, \bibinfo{pages}{12143} (\bibinfo{year}{2022}).

\bibitem{Kitai2020}
\bibinfo{author}{Kitai, K.} \emph{et~al.}
\newblock \bibinfo{journal}{\bibinfo{title}{Designing metamaterials with quantum annealing and factorization machines}}.
\newblock {\emph{\JournalTitle{Physical Review Research}}} \textbf{\bibinfo{volume}{2}}, \bibinfo{pages}{013319} (\bibinfo{year}{2020}).

\bibitem{Nawa2023}
\bibinfo{author}{Nawa, K.}, \bibinfo{author}{Suzuki, T.}, \bibinfo{author}{Masuda, K.}, \bibinfo{author}{Tanaka, S.} \& \bibinfo{author}{Miura, Y.}
\newblock \bibinfo{journal}{\bibinfo{title}{{Quantum Annealing Optimization Method for the Design of Barrier Materials in Magnetic Tunnel Junctions}}}.
\newblock {\emph{\JournalTitle{Physical Review Applied}}} \textbf{\bibinfo{volume}{20}}, \bibinfo{pages}{024044} (\bibinfo{year}{2023}).

\bibitem{Sampei2023}
\bibinfo{author}{Sampei, H.} \emph{et~al.}
\newblock \bibinfo{journal}{\bibinfo{title}{{Quantum Annealing Boosts Prediction of Multimolecular Adsorption on Solid Surfaces Avoiding Combinatorial Explosion}}}.
\newblock {\emph{\JournalTitle{Journal of the American Chemical Society}}} \textbf{\bibinfo{volume}{3}}, \bibinfo{pages}{991--996} (\bibinfo{year}{2023}).

\bibitem{Borrvall2003}
\bibinfo{author}{Borrvall, T.} \& \bibinfo{author}{Petersson, J.}
\newblock \bibinfo{journal}{\bibinfo{title}{{Topology optimization of fluids in Stokes flow}}}.
\newblock {\emph{\JournalTitle{International Journal for Numerical Methods in Fluids}}} \textbf{\bibinfo{volume}{41}}, \bibinfo{pages}{77--107} (\bibinfo{year}{2003}).

\bibitem{Challis2009}
\bibinfo{author}{Challis, V.~J.} \& \bibinfo{author}{Guest, J.~K.}
\newblock \bibinfo{journal}{\bibinfo{title}{{Level set topology optimization of fluids in Stokes flow}}}.
\newblock {\emph{\JournalTitle{International Journal for Numerical Methods in Engineering}}} \textbf{\bibinfo{volume}{79}}, \bibinfo{pages}{1284--1308} (\bibinfo{year}{2009}).

\bibitem{Duan2008}
\bibinfo{author}{Duan, X.}, \bibinfo{author}{Ma, Y.} \& \bibinfo{author}{Zhang, R.}
\newblock \bibinfo{journal}{\bibinfo{title}{Optimal shape control of fluid flow using variational level set method}}.
\newblock {\emph{\JournalTitle{Physics Letters A}}} \textbf{\bibinfo{volume}{372}}, \bibinfo{pages}{1374--1379} (\bibinfo{year}{2008}).

\bibitem{Zhou2008}
\bibinfo{author}{Zhou, S.} \& \bibinfo{author}{Li, Q.}
\newblock \bibinfo{journal}{\bibinfo{title}{{A variational level set method for the topology optimization of steady-state Navier--Stokes flow}}}.
\newblock {\emph{\JournalTitle{Journal of Computational Physics}}} \textbf{\bibinfo{volume}{227}}, \bibinfo{pages}{10178--10195} (\bibinfo{year}{2008}).

\bibitem{Yamada2010}
\bibinfo{author}{Yamada, T.}, \bibinfo{author}{Izui, K.}, \bibinfo{author}{Nishiwaki, S.} \& \bibinfo{author}{Takezawa, A.}
\newblock \bibinfo{journal}{\bibinfo{title}{A topology optimization method based on the level set method incorporating a fictitious interface energy}}.
\newblock {\emph{\JournalTitle{Computer Methods in Applied Mechanics and Engineering}}} \textbf{\bibinfo{volume}{199}}, \bibinfo{pages}{2876--2891} (\bibinfo{year}{2010}).

\bibitem{Olesen2006}
\bibinfo{author}{Olesen, L.~H.}, \bibinfo{author}{Okkels, F.} \& \bibinfo{author}{Bruus, H.}
\newblock \bibinfo{journal}{\bibinfo{title}{{A high-level programming-language implementation of topology optimization applied to steady-state Navier--Stokes flow}}}.
\newblock {\emph{\JournalTitle{International Journal for Numerical Methods in Engineering}}} \textbf{\bibinfo{volume}{65}}, \bibinfo{pages}{975--1001} (\bibinfo{year}{2006}).

\bibitem{Deng2011}
\bibinfo{author}{Deng, Y.}, \bibinfo{author}{Liu, Z.}, \bibinfo{author}{Zhang, P.}, \bibinfo{author}{Liu, Y.} \& \bibinfo{author}{Wu, Y.}
\newblock \bibinfo{journal}{\bibinfo{title}{{Topology optimization of unsteady incompressible Navier--Stokes flows}}}.
\newblock {\emph{\JournalTitle{Journal of Computational Physics}}} \textbf{\bibinfo{volume}{230}}, \bibinfo{pages}{6688--6708} (\bibinfo{year}{2011}).

\bibitem{Kondoh2012}
\bibinfo{author}{Kondoh, T.}, \bibinfo{author}{Matsumori, T.} \& \bibinfo{author}{Kawamoto, A.}
\newblock \bibinfo{journal}{\bibinfo{title}{Drag minimization and lift maximization in laminar flows via topology optimization employing simple objective function expressions based on body force integration}}.
\newblock {\emph{\JournalTitle{Structural and Multidisciplinary Optimization}}} \textbf{\bibinfo{volume}{45}}, \bibinfo{pages}{693--701} (\bibinfo{year}{2012}).

\bibitem{Gersborg2005}
\bibinfo{author}{Gersborg-Hansen, A.}, \bibinfo{author}{Sigmund, O.} \& \bibinfo{author}{Haber, R.~B.}
\newblock \bibinfo{journal}{\bibinfo{title}{Topology optimization of channel flow problems}}.
\newblock {\emph{\JournalTitle{Structural and Multidisciplinary Optimization}}} \textbf{\bibinfo{volume}{30}}, \bibinfo{pages}{181--192} (\bibinfo{year}{2005}).

\bibitem{Papoutsis2011}
\bibinfo{author}{Papoutsis-Kiachagias, E.}, \bibinfo{author}{Kontoleontos, E.}, \bibinfo{author}{Zymaris, A.}, \bibinfo{author}{Papadimitriou, D.} \& \bibinfo{author}{Giannakoglou, K.}
\newblock \bibinfo{title}{Constrained topology optimization for laminar and turbulent flows, including heat transfer}.
\newblock In \emph{\bibinfo{booktitle}{EUROGEN 2011 PROCEEDINGS --- Evolutionary and Deterministic Methods for Design, Optimization and Control with Applications to Industrial and Societal Problems}} (\bibinfo{year}{2011}).

\bibitem{Yoon2010}
\bibinfo{author}{Yoon, G.~H.}
\newblock \bibinfo{journal}{\bibinfo{title}{Topology optimization for stationary fluid--structure interaction problems using a new monolithic formulation}}.
\newblock {\emph{\JournalTitle{International Journal for Numerical Methods in Engineering}}} \textbf{\bibinfo{volume}{82}}, \bibinfo{pages}{591--616} (\bibinfo{year}{2010}).

\bibitem{Bendsoe2013}
\bibinfo{author}{Bends\o{}e, M.~P.} \& \bibinfo{author}{Sigmund, O.}
\newblock \emph{\bibinfo{title}{Topology optimization: theory, methods, and applications}} (\bibinfo{publisher}{Springer Science \& Business Media}, \bibinfo{year}{2013}).

\bibitem{Allaire2002}
\bibinfo{author}{Allaire, G.}, \bibinfo{author}{Jouve, F.} \& \bibinfo{author}{Toader, A.-M.}
\newblock \bibinfo{journal}{\bibinfo{title}{A level-set method for shape optimization}}.
\newblock {\emph{\JournalTitle{Comptes Rendus Mathematique}}} \textbf{\bibinfo{volume}{334}}, \bibinfo{pages}{1125--1130} (\bibinfo{year}{2002}).

\bibitem{Wang2003}
\bibinfo{author}{Wang, M.~Y.}, \bibinfo{author}{Wang, X.} \& \bibinfo{author}{Guo, D.}
\newblock \bibinfo{journal}{\bibinfo{title}{A level set method for structural topology optimization}}.
\newblock {\emph{\JournalTitle{Computer Methods in Applied Mechanics and Engineering}}} \textbf{\bibinfo{volume}{192}}, \bibinfo{pages}{227--246} (\bibinfo{year}{2003}).

\bibitem{Allaire2004}
\bibinfo{author}{Allaire, G.}, \bibinfo{author}{Jouve, F.} \& \bibinfo{author}{Toader, A.-M.}
\newblock \bibinfo{journal}{\bibinfo{title}{Structural optimization using sensitivity analysis and a level-set method}}.
\newblock {\emph{\JournalTitle{Journal of Computational Physics}}} \textbf{\bibinfo{volume}{194}}, \bibinfo{pages}{363--393} (\bibinfo{year}{2004}).

\bibitem{Talischi2012}
\bibinfo{author}{Talischi, C.}, \bibinfo{author}{Paulino, G.~H.}, \bibinfo{author}{Pereira, A.} \& \bibinfo{author}{Menezes, I.~F.}
\newblock \bibinfo{journal}{\bibinfo{title}{{PolyTop: a Matlab implementation of a general topology optimization framework using unstructured polygonal finite element meshes}}}.
\newblock {\emph{\JournalTitle{Structural and Multidisciplinary Optimization}}} \textbf{\bibinfo{volume}{45}}, \bibinfo{pages}{329--357} (\bibinfo{year}{2012}).

\bibitem{Pereira2016}
\bibinfo{author}{Pereira, A.}, \bibinfo{author}{Talischi, C.}, \bibinfo{author}{Paulino, G.~H.}, \bibinfo{author}{M.~Menezes, I.~F.} \& \bibinfo{author}{Carvalho, M.~S.}
\newblock \bibinfo{journal}{\bibinfo{title}{{Fluid flow topology optimization in PolyTop: stability and computational implementation}}}.
\newblock {\emph{\JournalTitle{Structural and Multidisciplinary Optimization}}} \textbf{\bibinfo{volume}{54}}, \bibinfo{pages}{1345--1364} (\bibinfo{year}{2016}).

\bibitem{Yaji2014}
\bibinfo{author}{Yaji, K.} \emph{et~al.}
\newblock \bibinfo{journal}{\bibinfo{title}{{Topology optimization using the lattice Boltzmann method incorporating level set boundary expressions}}}.
\newblock {\emph{\JournalTitle{Journal of Computational Physics}}} \textbf{\bibinfo{volume}{274}}, \bibinfo{pages}{158--181} (\bibinfo{year}{2014}).

\bibitem{Endo2024}
\bibinfo{author}{Endo, K.}, \bibinfo{author}{Matsuda, Y.}, \bibinfo{author}{Tanaka, S.} \& \bibinfo{author}{Muramatsu, M.}
\newblock \bibinfo{journal}{\bibinfo{title}{{Novel real number representations in Ising machines and performance evaluation: Combinatorial random number sum and constant division}}}.
\newblock {\emph{\JournalTitle{PLoS ONE}}} \textbf{\bibinfo{volume}{19}}, \bibinfo{pages}{e0304594} (\bibinfo{year}{2024}).

\bibitem{Fixstars}
\bibinfo{title}{{Fixstars Amplify Annealing Engine (AE)}}.
\newblock \bibinfo{note}{Available online: \url{https://amplify.fixstars.com/en/engine}}.

\bibitem{Key2024TopoFlow}
\bibinfo{author}{Key, F.}, \bibinfo{author}{Aoki, S.} \& \bibinfo{author}{Muramatsu, M.}
\newblock \bibinfo{title}{{EngiOptiQA/TopoFlow: v0.1.0}}, \doiprefix\url{10.5281/zenodo.14129614} (\bibinfo{year}{2024}).

\end{thebibliography}

\end{document}